\documentclass{aa}  

\usepackage{graphicx}

\usepackage{txfonts}
\usepackage{xcolor}

\begin{document} 

   \title{Contrasting evolutionary pathways of fast- and slow-rotating galaxies in the green valley}
   \titlerunning{Evolution of green valley fast- and slow-rotating galaxies}

   \author{S.~Zhou\inst{1},
          A.~Iovino \inst{1},
          M.~Longhetti \inst{1},
          F.~La Barbera \inst{2} and
          L.~Costantin \inst{3}
          } 

   \institute{INAF-Osservatorio Astronomico di Brera, via Brera 28, I-20121 Milano, Italy\\
              \email{shuang.zhou@inaf.it}
    \and{INAF - Osservatorio Astronomico di Capodimonte, Via Moiariello 16, I-80131 Napoli, Italy}  
    \and{Centro de Astrobiolog\'ia (CAB), CSIC-INTA, Ctra. de Ajalvir km 4, Torrej\'on de Ardoz, E-28850, Madrid, Spain}  
             }

   \date{Received XXX; accepted XXX}

  \abstract  
   {}
   {We present an investigation of the evolutionary pathways of green valley (GV) galaxies, defined as galaxies with a distance to the star-forming main sequence in the range $-1.3<\Delta_{\rm SFMS}<-0.5$, drawn from the SDSS-IV/MaNGA survey. Our goal is to examine the connection between a galaxy’s dynamical structure, specifically its stellar angular momentum, and its key physical properties, including gas-phase metallicity, star formation history (SFH), and chemical enrichment history (ChEH). By exploring these correlations, we aim to constrain the physical processes that govern  the evolution of green valley galaxies.}
   {We divided our sample into fast- and slow-rotating galaxies using a criterion motivated by the bimodal distribution of stellar spin and compared their integrated stellar and gas-phase metallicities.  Additionally, using a simple yet comprehensive galaxy chemical evolution model, optimised to fit for each galaxy the gas-phase metallicity estimates and the integrated stellar spectra, we reconstructed the galaxies’ past star formation and chemical enrichment histories. The derived SFHs and ChEHs, along with parameters governing the gas infall timescale and outflow strengths, offer valuable insights into the physical processes driving the evolution and quenching of our GV sample galaxies.}
   {We found that fast-rotating galaxies exhibit systematically higher metallicities in both the gas and stellar phases compared to slow-rotating galaxies. However, in the gas phase, the difference is significant only at the low-mass end, while the stellar metallicity offset is observed consistently across the full stellar mass range. Our modelling framework yields simple yet physically motivated explanations for these trends. At low stellar masses, the model predicts similar gas-infall and star-formation timescales for fast- and slow-rotating galaxies, but the stronger outflows inferred for the slower population substantially reduce their chemical content in both gas and stars. At high masses, slow-rotating galaxies show shorter gas-infall and star-formation timescales; the combination of reduced pristine gas inflow and more efficient gas removal produces gas-phase metallicities comparable to fast-rotating galaxies but systematically lower stellar metallicities.}
   {The systematic differences in stellar and gas-phase metallicity, as well as in model-inferred gas-accretion and outflow parameters, highlight the contrasting properties of fast- and slow-rotating galaxies in our GV sample. Interpreting these differences within our chemical-evolution framework and combining evidence from theoretical studies suggests distinct past evolution. Slow-rotating galaxies likely experienced more merger events, usually associated with strong gas removal processes, leading to their systematically lower metallicities. At low masses, stronger supernova-driven outflows reduce their chemical content while leaving star-formation timescales similar to fast-rotating galaxies. At high masses, merger-triggered AGN feedback may rapidly deplete and suppress gas infall, producing the shorter star-formation timescales seen in slow-rotating galaxies. Alternative environmental and assembly-driven scenarios are also discussed.
}


   \maketitle

\section{Introduction}
Since the original discovery by Edwin Hubble, a century ago, of the existence of different classes of morphological types \citep{Hubble1926ApJ}, galaxies have been broadly divided into spiral and elliptical galaxies (e.g., \citealt{deVaucouleurs1959}).  Astronomers also noticed very early that ellipticals are generally red in colour, while spirals appear to be mostly blue (e.g., \citealt{Holmberg1958}). 
Yet such bimodality was not fully explored until the advent of large-scale surveys such as The Sloan Digital
Sky Survey (SDSS, \citealt{York2000}). The unprecedented amount of galaxies observed by SDSS shows a sharp bimodality in colour, forming a so-called blue cloud and red sequence on the colour-magnitude diagram \citep{Strateva2001, Baldry2004}. Subsequent analysis reveals that the blue cloud generally consists of star-forming galaxies, positioned along a star-forming main sequence (SFMS) on the star formation rate-stellar mass diagram \citep{Brinchmann2004,Noeske2007}, while the red sequence is occupied by quenched, passively evolving galaxies \citep{Bell2004,Schiminovich2007}. The transition of galaxies from the blue cloud to the red sequence has been a topic of intense interest in extragalactic studies. Falling between the red sequence and the blue cloud, the so-called green valley galaxies have also received much attention, as they are believed to be transitioning from one sequence to the other and therefore may shed light on when and why galaxies leave the SFMS (e.g., \citealt{Mendez2011,Schawinski2014, Salim2014}).

A number of investigations have found that the changing of colour, as well as of the star formation state, in a galaxy is accompanied by the variation of some structural parameters (e.g., 
\citealt{Kauffmann2003,Cameron2009,Wuyts2011,Bell2012,Lang2014,Whitaker2017}). More detailed morphological properties could not be taken into account until reliable classifications became available, through the 
Galaxy Zoo project \citep{Lintott2011}. Using these morphological classifications, \cite{Schawinski2014} report that two distinct populations in fact occupy the green valley. Late-type green valley galaxies proceed towards the red sequence due to the shutdown of their gas supply and the exhaustion of their remaining gas over several Gyrs. In contrast, early-type green valley galaxies undergo rapid quenching, characterised by the abrupt destruction of their gas reservoirs, accompanied by a morphological transformation from a disk to a spheroid. Such a difference indicates distinct physical processes driving the evolution of different types of galaxies from the blue cloud to the red sequence.

The development of integral field spectroscopy in the past two decades has enabled us to obtain precise spatial kinematic properties of galaxies, complementing the structural and morphological indicators inferred from broadband images, thus providing more decisive constraints on assessing the galaxy's intrinsic morphology (see the extensive review by \citealt{Cappellari2016ARA}). Using the kinematic information, galaxies can be classified into slow rotators, which show little to no rotation in their stellar velocity maps, and fast rotators, which manifest clear rotational features (e.g., \citealt{Emsellem2007, Cappellari2007, Cappellari2011, Krajnovic2011}). The bimodality in the angular momentum of the stellar components in galaxies is found to be fundamental \citep{Cappellari2016ARA, Graham2018, vandeSande2021}, and appears in galaxies at different star formation levels and in different environments \citep{Wang2024Nat}. Similar to the findings in morphology, Wang et al. (2024) suggest that galaxies of different kinematic states can undergo divergent evolutionary pathways when they quench their star formation. By finding a lack of dependence of the stellar metallicity on the star formation rate in slow-rotating galaxies, the authors speculate that these galaxies may have quenched through merger-induced events associated with strong gas outflows, while fast-rotating galaxies evolve secularly.

In this work we present a detailed analysis combining evidence from the stellar and gas phases to investigate the quenching processes affecting galaxies in the green valley. We use data from the Mapping Nearby Galaxies at Apache Point Observatory (MaNGA; \citealt{Bundy2015}) survey, which provides high-quality integral field unit (IFU) data enabling accurate kinematic classifications. The large sample allows us to select a substantial number of green valley galaxies in transition. Building on \cite{Zhou2022}, we show that combining stellar and gas-phase diagnostics gives a comprehensive view of star-formation and chemical-enrichment histories. We first derive the global stellar and gas properties of our sample to identify systematic differences linked to their kinematic states. We then apply a semi-analytic spectral fitting approach, fitting full chemical-evolution models directly to the observed spectra and gas-phase metallicities rather than relying on simple average ages or metallicities. This method traces mass assembly and enrichment histories in detail and constrains key parameters related to gas inflows and outflows, directly linking them to the physical processes driving evolution. By combining the large dataset with advanced modelling tools, we aim to clarify the processes that govern the evolution of green valley galaxies.

Our paper is organised as follows. We present the data, including the IFU spectra from MaNGA and auxiliary data products in \S\ref{sec:data} and define the sample used in this work in \S\ref{sec:sample}. We then present a galaxy evolution model to explain the observables, and discuss how we fit the model to the observed data in \S\ref{sec:analysis}. The inferred evolutionary paths and parameters from the best-fit models are presented in \S\ref{sec:results}, and the physics we can infer from the results are discussed in \S\ref{sec:discussion}. We finally summarise our key findings in \S\ref{sec:summary}. Throughout this work we use a standard $\Lambda$CDM cosmology with $\Omega_{\Lambda}=0.7$, 
$\Omega_{\rm M}=0.3$ and $H_0$=70km s$^{-1}$ Mpc$^{-1}$, which are rounded off WMAP values \citep{Bennett2003}. 

\section{Data}
\label{sec:data}

\subsection{MaNGA}
As part of the fourth generation of SDSS (SDSS-IV, \citealt{Blanton2017}), the MaNGA collaboration obtained and made public spatially-resolved spectroscopy data for more than 10,000 nearby (redshift $0.01<z<0.15$) galaxies \citep{Yana2016}. The MaNGA galaxy sample has an almost flat stellar mass distribution covering the range $5\times10^8 h^{-2}{M}_{\odot} \leq M_*\leq 3 \times 10^{11} h^{-2}{M}_{\odot}$ \citep{Wake2017}. 
Spatially-resolved spectroscopy was taken to cover at least out to 1.5 effective radii (R$_{\rm e}$) for all the target galaxies \citep{Law2015}, providing intermediate spectral resolution ($R\sim2000$, \citealt{Drory2015}) spectral data within the wavelength range $3600-10300$ {\AA} using the two dual-channel BOSS spectrographs \citep{smee2013} mounted on the Sloan 2.5\,m telescope \citep{Gunn2006}. The raw MaNGA data were reduced using the dedicated  
Data Reduction Pipeline (DRP; \citealt{Law2016}), which produces spectra with flux calibration better than $5\%$ throughout the covered wavelength range \citep{Yanb2016}. In addition, the MaNGA team distributes various useful data products, including spatially resolved stellar kinematics, emission-line properties, and spectral indices, obtained by analysing the observed spectra with the Data Analysis Pipeline (DAP; \citealt{Westfall2019, Belfiore2019}). All these data are
released as part of the final data release of SDSS-IV \citep[DR17\footnote{\url{https://www.sdss.org/dr17/manga/}};][]{SDSSDR17}. 
Further extensive analysis of the data carried out by the MaNGA collaboration resulted in the release of different value-added catalogues (VACs)\footnote{\url{https://www.sdss4.org/dr17/data_access/value-added-catalogs/}}. 

\subsection{The green valley}
\begin{figure}
    \centering
    \includegraphics[width=0.45\textwidth]{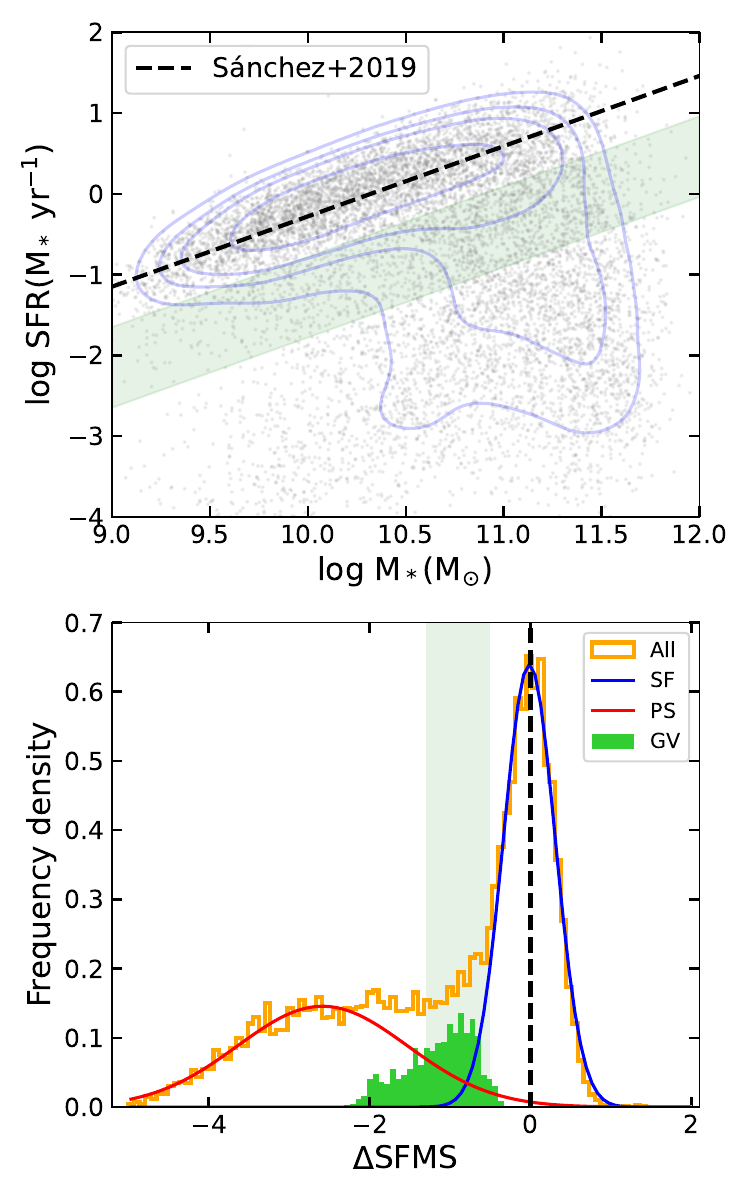}\\
     \caption{Top: the star formation rate as a function of stellar mass for MaNGA galaxies. Grey dots are individual MaNGA galaxies, with contours enclosing 
     20\%, 40\%, 60\% and 80\% of the data points. The solid line shows the SFMS obtained from \cite{Sanchez2019}.
     Bottom: distribution of $\Delta_{\rm SFMS}
     $ for MaNGA galaxies. Orange histogram shows the entire sample, with blue and red lines showing the possible distribution of star-forming and passive galaxies, obtained by fitting a Gaussian function to the histogram with -0.5<$\Delta_{\rm SFMS}$ and $\Delta_{\rm SFMS}$<-2.5, respectively. The green histogram shows the possible distribution of green valley galaxies by subtracting the contribution of star-forming and passive galaxies from the whole distribution. In both panels, the shaded region indicates the green valley used in this work defined as the region of -1.3<$\Delta_{\rm SFMS}$<-0.5.
     }
     \label{fig:SFMS}
\end{figure}

To quantify the star formation status of MaNGA galaxies, we used the global star formation rate (SFR) provided by the Pipe3D value-added catalogue \citep{Sanchez2016, Sanchez2022pipe3d}.  In the catalogue, the SFRs are calculated from the
dust-corrected H$\alpha$ line fluxes obtained by integrating the H$\alpha$ fluxes across all the available spaxels of each galaxy. As shown by \cite{Sanchez2022pipe3d}, this value well represents the star formation activity in each MaNGA galaxy. It can be seen from the top panel of Fig.~\ref{fig:SFMS} that, using this quantity, the MaNGA galaxies display the well-known star-forming main sequence (SFMS) on the SFR-M$_*$ plane. The star formation state of galaxies can then be quantified by their distance from the SFMS at a given stellar mass:
\begin{equation}
    \Delta_{\rm SFMS}\equiv \log {\rm SFR}-\log {\rm SFMS}(M_*),
\end{equation}
where we use the SFMS defined in \cite{Sanchez2019} (shown as a dashed line in the top panel of Fig.~\ref{fig:SFMS}) as 
\begin{equation}
    \log {\rm SFMS}(M_*) = -8.96 + 0.87\times\log M_*.
\end{equation}

We then define the GV region using the distribution of $\Delta_{\rm SFMS}$ of all MaNGA galaxies shown in the bottom panel of Fig.~\ref{fig:SFMS}. For $\Delta_{\rm SFMS} > -0.5$, we fit a Gaussian function to the distribution (blue line) and obtain a best-fit dispersion of $\sigma = 0.37$, which provides an excellent description of the profile. This component is dominated by star-forming (SF) galaxies and traces the primary locus of the star-forming main sequence. On the low $\Delta_{\rm SFMS}$ end — primarily populated by passive (PS) galaxies — the distribution is also approximately Gaussian (red line), though with a broader spread due to larger uncertainties in SFR estimates at these low values. By subtracting the two Gaussians representing the SF and PS populations, we isolate the residual distribution (green histogram), which likely corresponds to green valley galaxies. This population significantly overlaps with both SF and PS galaxies at the high and low $\Delta_{\rm SFMS}$ ends. To obtain a cleaner sample, we define green valley galaxies as those within the range $-1.3 < \Delta_{\rm SFMS} < -0.5$, as indicated by the green shaded region in Fig.~\ref{fig:SFMS}. As seen from Fig.~\ref{fig:SFMS}, this selection reaches a good balance between including as many green valley galaxies as possible while minimising contamination from the star-forming and passive populations.

We acknowledge that alternative choices exist for estimating the global SFR for these MaNGA galaxies. For example, \citet{Wang2024Nat} obtained SFRs for MaNGA galaxies by cross-matching with the GALEX-SDSS-WISE Legacy Catalogue \citep{Salim2016GSWLC}, where SFRs are obtained through fitting the UV-to-mid-IR spectral energy distribution (SED). However, as shown in their Fig. 3, at fixed SED-based SFRs, galaxies exhibit a non-negligible scatter in their H$\alpha$ fluxes, and therefore in H$\alpha$-based SFR estimates, that correlates with their kinematic state. Such scatter is particularly evident for green valley galaxies. As noted by \citet{Wang2024Nat}, this discrepancy arises partly because H$\alpha$ traces star formation over shorter timescales, whereas SED fitting averages star formation over more extended periods. Moreover, SED fitting typically relies on assumed parametric forms for the star formation history \citep{Salim2016GSWLC}, which inherently ties the inferred recent SFR to the galaxy’s past evolution. This coupling can introduce systematic biases in the selection process between fast and slow-rotating galaxies if their evolutionary histories differ. Since our goal is to identify green valley galaxies that are likely undergoing rapid transitions in star formation activities, H$\alpha$-based SFRs are better suited to reflect their instantaneous star-forming state and are less sensitive to assumptions about long-term star formation histories. By exploring green-valley selections constructed using different SFR indicators, we find that although they produce slightly different samples, the resulting galaxies exhibit consistent global properties and lead to the same qualitative evolutionary picture. After confirming this robustness, we decide to proceed with the H$\alpha$-based SFRs throughout this work.

\subsection{Kinematic measurements}
For each MaNGA galaxy, \cite{Wang2023APJL} provide the kinematic information that we will use in this paper. We briefly summarise here how this information is derived in their work.

The stellar spin parameter $\lambda_{Re}$ is a widely used parameter to characterise the angular momentum of a galaxy \citep{Emsellem2007, Cappellari2007, Cappellari2011}. In \cite{Wang2023APJL}, this parameter is measured for 9,793 MaNGA galaxies using a similar procedure as in \cite{Graham2018}. To begin with, stellar kinematic maps, with the Voronoi spatial binning to signal-to-noise ratio of $\sim10$, are taken from MaNGA DAP. In each pixel of these maps,  the stellar kinematics is characterised by the mean stellar velocity $V$ and velocity dispersion $\sigma$. After correcting the mean stellar velocity for the systemic velocity and the velocity dispersion for instrument broadening, the spin parameter is computed using \citep{Emsellem2007}: 
\begin{equation}
\lambda_{R}\,\equiv \, \frac{\left< R\,|V| \right>}{\left< R\, \sqrt{V^2+\sigma ^2} \right>}\,=\,\frac{\sum_{n=1}^{N} F_n R_n |V_n|}{\sum_{n=1}^{N} F_n R_n \sqrt{V_n^2+\sigma _n ^2}}
\end{equation}
where $F_n$, $V_n$  and $\sigma_n$  are the flux, mean velocity and velocity dispersion of the nth pixel.

The calculation is made within the elliptical half-light radius $R_{\rm e}$, provided by \cite{Zhu2023}. To obtain this quantity \cite{Zhu2023} fits the SDSS r-band photometry from NASA-Sloan Atlas\footnote{http://www.nsatlas.org} (NSA; \citealt{Blanton2011}) to obtain for each galaxy an isophote containing half of the total luminosity. The ellipticity $\varepsilon$ of a galaxy is then calculated as the second moment of the light distribution inside the half-light isophote :
\begin{equation}
(1-\varepsilon)^2\,=\,q^{\prime 2}\,=\,\frac{\left< y^2 \right>}{\left< x^2 \right>}\,=\,\frac{\sum_{k=1}^{P} F_k y_k^2}{\sum_{k=1}^{P} F_k x_k^2}
\end{equation}
,where $F_k$ and $(x_k,y_k)$ are the flux and coordinates of the $k$th pixel.

\begin{figure}
    \centering
    \includegraphics[width=0.55\textwidth]{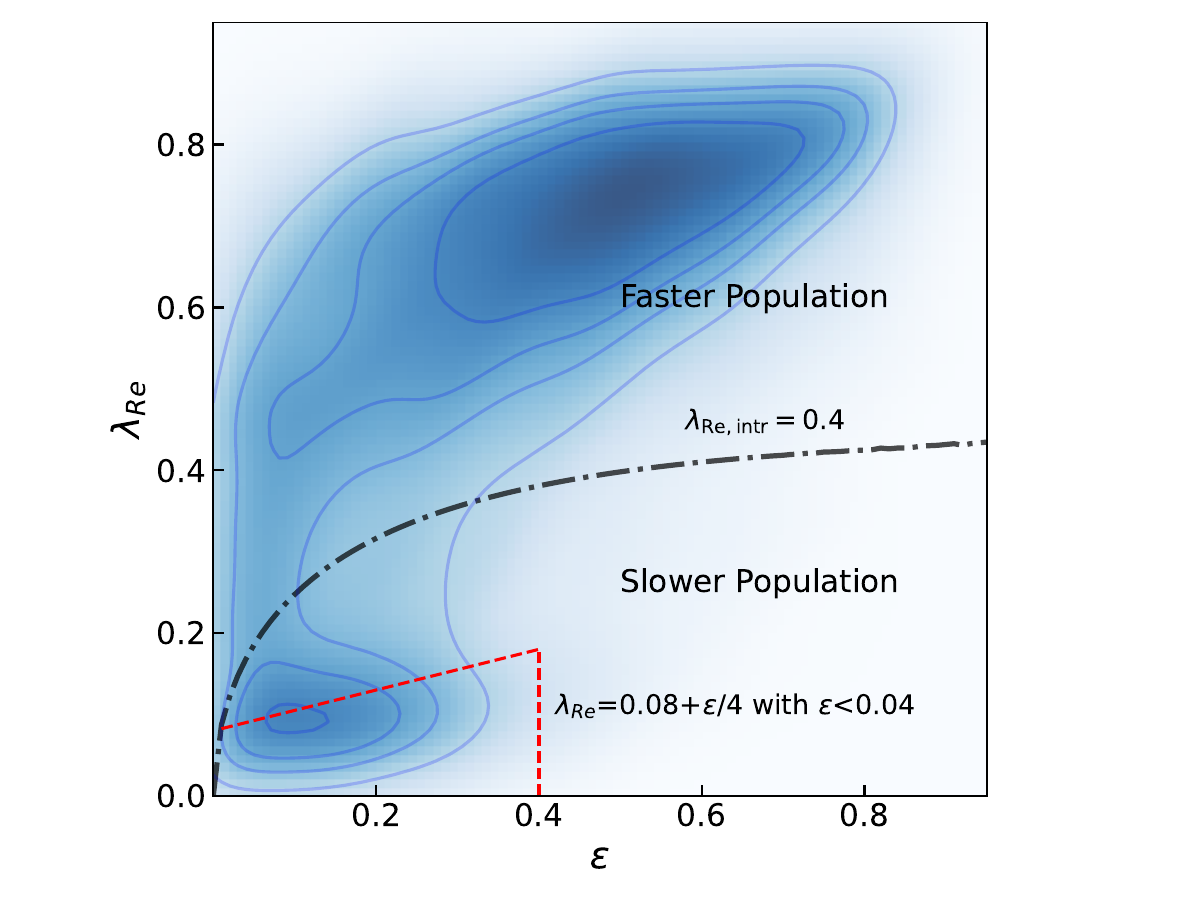}\\
     \caption{The  spin parameter $\lambda_{Re}$ of MaNGA galaxies  as a function of their ellipticity   $\varepsilon$. Colours indicate the density of the sample galaxies, with contours enclosing 
     20\%, 40\%, 60\% and 80\% of the data points. The black dash-dotted line represents the theoretically predicted locus for an axisymmetric galaxy with $\lambda_{\rm Re, intr}=0.4$ (assuming anisotropy $\delta=0.7\times\varepsilon_{\rm intr}$) viewed over all inclinations, as proposed by \cite{Wang2023APJL}. Throughout this work, galaxies above and below this curve are defined as the faster and slower populations, respectively. For comparison, the classical fast/slow-rotator division from \cite{Cappellari2016ARA} is also shown as a red dashed curve.} 
     \label{fig:Ki_dis}
\end{figure}

During the calculation, \cite{Wang2023APJL} apply
seeing-related corrections to the observed $\lambda_{\rm Re}$ values following the procedure described in \citet{Graham2018}. In addition, they provide quality flags that indicate the goodness of the spin parameter as obtained from the above method. We use only the galaxies flagged with `clean'=y, which yields a clean sample of 8,639 galaxies with reliable measurements in $\lambda_{Re}$ and $\varepsilon$. In Fig.~\ref{fig:Ki_dis} we plot the spin parameter $\lambda_{Re}$ of these galaxies as a function of their ellipticity $\varepsilon$. It is clear that the galaxies are mainly distributed in two separate clouds. The fast-rotating population generally exhibits high spin parameters, showing strongly rotation-supported behaviour, while the slow-rotating population occupies the bottom region of the $\lambda_{Re}$  - $\varepsilon$ space. As extensively discussed by \citet{Wang2024Nat}, such bimodality is observed ubiquitously across a wide range of star-formation levels and environments, and an intrinsic (i.e., measured for an edge-on orientation) stellar spin  of $\lambda_{\rm Re, intr}=0.4$ empirically separates galaxies with fast and slow rotation. The dash-dotted curve shows the theoretically predicted locus for an axisymmetric galaxy with $\lambda_{\rm Re, intr}=0.4$ (assuming an anisotropy $\delta=0.7\times\varepsilon_{\rm intr}$), viewed over all possible inclinations. Following \citet{Wang2023APJL}, we adopt this curve to define the populations of faster (above the curve) and slower (below the curve) rotation, and we refer to them hereafter as the faster and slower populations wherever relevant. This division yields 6726 and 1913 galaxies in the faster and slower populations, respectively, within the MaNGA sample.

We note that the terminology adopted here, i.e., fast-rotating and slow-rotating galaxies, is intentionally distinct from the classical fast and slow rotators. For comparison, we show the division line for fast/slow rotators from \cite{Cappellari2016ARA} as a red dashed curve in  Fig.~\ref{fig:Ki_dis}.  As is immediately evident from Fig.~\ref{fig:Ki_dis}, the classical division line passes through the density peak of the slow-rotating population in our sample and therefore does not provide an optimal separation. As discussed in \cite{Wang2023APJL}, this difference arises mainly from variations in sample selection. Classical analyses \citep[e.g.,][]{Emsellem2011,Cappellari2016ARA} are typically limited to early-type galaxies, whereas the definition adopted here is tied to the empirical bimodality of intrinsic stellar angular momentum observed ubiquitously across diverse galaxy populations. Although \cite{Wang2023APJL} report that their definition differs only slightly from the classical scheme, we retain their nomenclature to ensure consistency and avoid confusion. For this reason, we refrain from using the classical terms “fast rotator” and “slow rotator” throughout this work and advise readers to be mindful of the distinction. Nevertheless, we verify that adopting the standard fast/slow-rotator classification does not affect our main results.

\section{Sample}
\label{sec:sample}
\subsection{Sample definition}
\label{ssec:sample}
We begin by selecting fast and slow-rotating  galaxies that satisfy our green valley criterion of $-1.3 < \Delta_{\rm SFMS} < -0.5$, resulting in a sample of 1302 fast-rotating and 187 slow-rotating galaxies. We further limit our analysis to the stellar mass range $10^{9.5}<M_*/{\rm M}_{\odot}<10^{11.5}$, following the suggestion by \cite{Wang2023APJL} that kinematic information becomes unreliable for the least massive galaxies with stellar mass below $10^{9.5}{\rm M}_{\odot}$, and considering that there are not many galaxies available above $10^{11.5}{\rm M}_{\odot}$ . This mass cut yields a final green valley sample of 1014 and 143 galaxies in the faster and slower populations. We divide the GV sample into stellar-mass bins of width 0.5 dex and report the number of galaxies in each bin in Table~\ref{tab:bins}. Given the similar stellar masses of galaxies within each selected bin, any observed differences in other properties between the two kinematic classes may reflect intrinsic differences in the evolutionary pathways of the faster and slower populations. 

In parentheses next to the galaxy counts in Table~\ref{tab:bins}, we also report the fraction of green valley fast- and slow-rotating galaxies relative to their respective total populations. Notably, the green valley fraction is systematically higher among the faster population compared to the slower population: a natural interpretation of this trend is that the quenching timescale differs between the two kinematic classes, suggesting a significant divergence in their evolutionary pathways. In other words, the slower population may experience a more rapid transition from the star-forming sequence to the quiescent red cloud, resulting in a lower likelihood of being observed in the intermediate green valley phase. Moreover, the most pronounced difference in green valley fractions appears among the most massive galaxies in our sample, while the disparity becomes marginal at lower masses. If the aforementioned interpretation holds, such an intriguing difference implies that the transition timescale disparity between the faster and slower populations may be mass-dependent.
Alternatively, the apparent mass dependence could also arise from the incomplete sampling of galaxies at different stellar masses in MaNGA, which may bias the relative fractions of the faster and slower populations and produce an artificial trend.
We will revisit this point in the following analysis.

\begin{table}
	\centering
	\caption{Number of green valley  fast- and slow-rotating galaxies in each stellar mass bin.}
	\label{tab:bins}
        \setlength{\tabcolsep}{4.2pt}
	\begin{tabular}{lcccc}
		\hline
		log($M_*/{\rm M}_{\odot})$ & 9.5-10.0 & 10.0-10.5 & 10.5-11.0& 11.0-11.5\\ 
		\hline
		Faster Pop. & 129 (11\%) & 194 (11\%)& 341 (17\%)& 350 (24\%)\\
		Slower Pop. & 26 (9\%) & 29 (11\%)& 34 (12\%)& 54 (7\%)\\
		\hline
	\end{tabular}
    \tablefoot{ The values in parentheses indicate their corresponding fractions relative to the total faster and slower populations, respectively.}

\end{table}

\begin{figure*}
    \centering
    \includegraphics[width=0.95\textwidth]{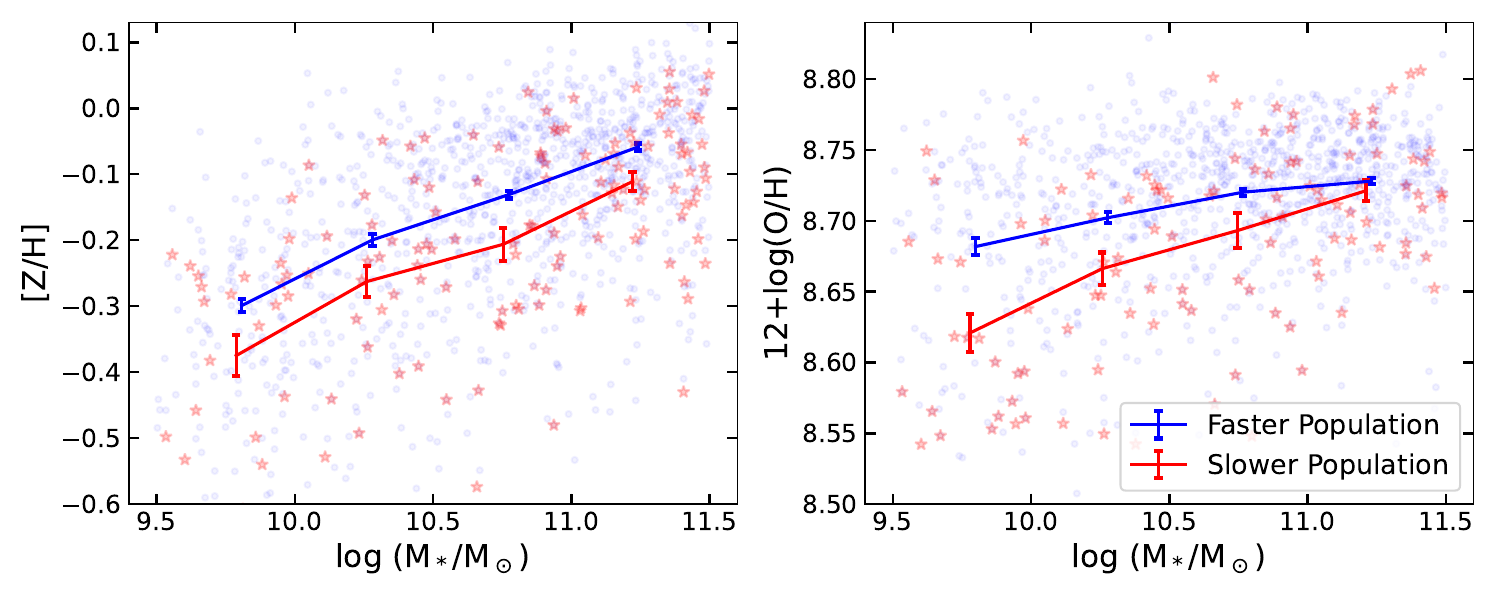}
     \caption{The stellar (left) and gas phase (right) metallicities as a function of stellar mass for our sample galaxies. In each panel, fast-rotating galaxies are shown in blue, while slow-rotating galaxies are shown in red.  The solid lines indicate the mean relation, with error bars representing the standard deviation of 1,000 bootstrap resamplings.
}
     \label{fig:pip3d_metal}
\end{figure*}

\subsection{Spectral stacking}
 The raw MaNGA spectra, provided by the MaNGA DRP for individual pixels, have relatively low signal-to-noise ratios (S/N), particularly in the outer regions of galaxies (typically $\sim 4$–8~{\AA}$^{-1}$; \citealt{Law2016}). Such low S/N limits the feasibility of reliably constraining chemical enrichment histories at the spaxel level \citep[e.g.][]{Zhou2022}. To mitigate this limitation, we adopt an approach of spatially co-adding spectra within 1R${\rm _e}$ of each galaxy. This stacking procedure yields a higher S/N integrated spectrum representative of the galaxy's global property,  allowing us to carefully model the stellar population and chemical evolution of each galaxy as a whole. We acknowledge that the physical scale encompassed by 1R${\rm _e}$ varies from galaxy to galaxy. However, after examining the R${\rm _e}$ distributions of the faster and slower populations in our green valley sample, we find that both populations follow a similar distribution, with a median R${\rm _e}$ of approximately 5 arcseconds and a standard deviation of about 2.8 arcseconds, showing no significant difference between them.
 This ensures that the properties compared are derived from regions of comparable physical extent in both the faster and slower populations.

The stacking procedure is similar to that described in \cite{Zhou2019}, and readers are referred to that paper for full details. In short, the spectra from individual pixels are shifted to the rest frame using the velocity field provided by the MaNGA DAP and then co-added to produce a stacked spectrum. During the stacking process, the covariance between spaxels is accounted for using the correction term provided by \cite{Westfall2019}. This procedure yields coadded spectra with a typical final S/N of 70 per {\AA}, which is well-suited for deriving reliable estimates of galaxy SFHs and ChEHs.

\subsection{Global galaxy properties}
\label{ssec:global}

Before analysing the full evolutionary histories of the galaxies, we first quickly
examine their average stellar properties within 1Re using the co-added 
spectra. This is done by applying the spectral fitting tool pPXF \citep{Cappellari2017} to the 
spectra stacked within 1Re. The fitting procedure follows the 
methodology described in \citet{Zhou2025}, to which we refer the reader for 
further details.

Briefly, pPXF is applied to the co-added spectra to 
 recover the line-of-sight velocity dispersions and model the stellar continuum. In the fitting, we employ the E-MILES\footnote{\label{foot:emiles}\url{http://miles.iac.es/}} stellar population synthesis models \citep{Vazdekis2016}, which are updated versions of the MILES models originally presented by \citet{Vazdekis2010}. Specifically, we adopt models computed with a Chabrier IMF \citep{Chabrier2003} and Padova2000 isochrones \citep{Girardi2000Padova}. These models span stellar ages from 0.063 to 17.78 Gyr and metallicity from 0.0001 to 0.03. The simple stellar populations (SSPs) are computed to cover the wavelength range
from 1680.2 {\AA} to $5{\rm \mu m}$, with a uniformly high resolution of 2.51 {\AA} (FWHM)
over the range from 3540 {\AA} to 8950 {\AA}. Our MaNGA galaxy spectra cover the wavelength range 3600–10300 {\AA} and have a median redshift of 0.03.
Since the main part of their covered restframe wavelength range falls within the high-resolution range of the E-MILES models (3540 - 8950 {\AA}), and wavelengths beyond 8950 {\AA} are strongly affected by telluric and sky-line residuals, we decided to limit our analysis to this restricted range. From the pPXF fitting results, we derive a 
light-weighted stellar metallicity that serves as the characteristic value for 
each stacked spectrum. These metallicities are plotted as a function of stellar mass in the left panel of Fig.~\ref{fig:pip3d_metal}.

The best-fit continuum is then subtracted from the stacked spectrum to obtain a residual spectrum, from which we measured the fluxes in the four emission lines \hbox{[O\,{\sc iii}]}$\lambda$5007, \hbox{[N\,{\sc ii}]}$\lambda$6584, H${\alpha}$, and H${\beta}$. These line fluxes are converted into gas-phase metallicities using the O3N2 calibrator of \cite{Curti2017}. The gas-phase metallicities are then used to represent the galaxy’s characteristic gas-phase metallicity and are plotted as a function of stellar mass in the right panel of Fig.~\ref{fig:pip3d_metal}. We verify that the adoption of different calibrations for the gas phase-metallicities will lead to systematic changes in the absolute values derived \citep[e.g.,][]{Kewley2008}, but all the global trends observed in this work remain unchanged. 

From Fig.~\ref{fig:pip3d_metal}, it is notable that the faster and slower populations exhibit distinct chemical compositions. In the stellar phase (left panel), the faster population consistently show higher metallicities across the entire stellar mass range explored in this study. In the gas phase (right panel), however, the slower population display a significant metallicity deficiency compared to the faster population at the low-mass end. This discrepancy diminishes with increasing mass, such that for the most massive galaxies in our sample, the two populations exhibit nearly identical gas-phase metallicities.

As a sanity check, we also derive characteristic stellar and gas-phase metallicities in the same 1Re region for each galaxy from the Pipe3D \citep{Sanchez2022pipe3d} catalogue by computing the median of the spaxel-based measurements within 1Re. As expected, these values differ slightly from those obtained through our pPXF analysis, reflecting the different methodologies and weighting schemes. However, the overall trends discussed above remain unchanged. In addition, although we present light-weighted metallicities, to ease comparison with \cite{Wang2024Nat}, who
adopt light-weighted measurements from the Pipe3D catalogue, we have verified that the mass-weighted metallicities exhibit similar trends, indicating that these findings are robust against the choice of measurement approach.

It is worth noting that the stellar metallicity difference between the two populations observed in our green valley sample is not present in the green valley sample defined by \citet{Wang2024Nat}. This discrepancy is not surprising. Our green valley selection is offset by approximately 0.2 dex toward lower SFRs compared to that of \citet{Wang2024Nat}. Given the similar widths of the Gaussian distributions of star-forming galaxies in both studies ($\sigma_{\rm \Delta SFMS}\sim0.37$ for both H$\alpha$- and SED-based SFRs used in the two works), our criterion places galaxies farther away from the star-forming sequence. In their Fig. 4, \citet{Wang2024Nat} show that the metallicity differences between the faster and slower populations only become significant as galaxies become sufficiently passive. Specifically, although no such difference is observed in their green valley sample, the slower population in their passive sample show lower stellar metallicities than the faster population, which is fully consistent with what we find in our more {\it passive} green valley sample. Our selection thus achieves a favourable balance: it identifies green valley galaxies that have evolved sufficiently from the star-forming sequence for differences in stellar population properties between the two populations to become evident, yet not so evolved that measurements of gas-phase metallicities (by means of their nebular emission) and the reconstruction of their past evolutionary path (by means of our modelling) become infeasible. 

\section{Analysis}
\label{sec:analysis}
In this section, we present a simple yet flexible chemical evolution model to help interpret the observed differences between fast- and slow-rotating green-valley galaxies. By fitting this model directly to the galaxy spectra and gas-phase metallicities, we can gain insights into the key physical processes and scenarios that may be driving the observed trends.

\subsection{Chemical evolution model}
\label{subsec:gas_metal_measure}
In the current picture of galaxy evolution \citep[e.g.][]{Mo1998}, cool gas accretes onto dark matter halos to fuel star formation, while massive stars enrich the interstellar medium with metals. Feedback from supernovae and AGNs can expel gas, suppress star formation, and halt further chemical enrichment.
 Combining all these ingredients, we can model the gas mass evolution of a galaxy through the following equation:
\begin{equation}
\label{eq:massevo}
\dot{M}_{\rm g}(t)=\dot{M}_{\rm in}(t)-\psi(t)+\dot{M}_{\rm re}(t)-\dot{M}_{\rm out}(t).
\end{equation}
The first term on the right-hand side characterises the gas falling into the galaxy. In the simplest case, we can assume that the gas infall process can be described with an exponentially decaying function that reads:
\begin{equation}
\dot{M}_{\rm in}(t)=A e^{-(t-t_0)/\tau}\ (t > t_0),
\end{equation}
where we have two free parameters: $t_0$ specifies the starting time of the gas infall, with $\tau$ characterising the timescale of this process.

The second term of Eq.~\ref{eq:massevo} represents the star formation process in the galaxy. For simplicity, we adopt a linear Schmidt law \citep{Schmidt1959} with 
 \begin{equation}
\psi(t)=S\times M_{\rm g}(t),
\end{equation}
to model such process, with $S$ being the star-formation efficiency and $M_{\rm g}$ being the mass of the galaxy's gas content. Many studies have shown that star formation efficiency can be largely dictated by measurable local properties 
\citep{Leroy2008,Shi2011}. In this work, we adopt the extended Schmidt law proposed by \citet{Shi2011}, in which the SFE can be easily estimated from the stellar mass surface density ($\Sigma_{*}$) via 
\begin{equation}
\label{eq:sfe}
    S (yr^{-1})=10^{-10.28\pm0.08} \left( \frac{\Sigma_*}{M_{\odot} pc^{-2}} \right)^{0.48}.
\end{equation}
To use the relation, we obtain an approximate average value for the stellar mass surface density for our sample galaxies from the values of mass and effective radius provided by the NASA-Sloan Atlas with $\Sigma_*=2\times M_*/(\pi R_e^2)$. In principle, the characteristic stellar surface density evolves over cosmic time, as galaxies grow in both stellar mass and physical size. However, as discussed in \citet{Zhou2022}, incorporating the estimated mass and size evolution of MaNGA galaxies from \citet{Peterken2020} results in less than a 10\% variation in the predicted SFE over the past 10 Gyr. The test demonstrates that this modest time evolution time-varying SFE has a minimal impact on the inferred star formation and chemical enrichment histories, resulting in negligible differences in the overall trends. Consequently, we adopt a fixed SFE in our modelling framework to reduce degeneracy and computational complexity without compromising the accuracy of our results.

The third term of Eq.~\ref{eq:massevo} indicates the return of gas from dying stars. Following canonical assumptions \citep{Spitoni2017, Zhou2022}, we adopt a constant mass return fraction of $R=0.3$, which means that around 30\% of the stellar mass formed in each generation can be returned to the ISM. We don’t account for any delay after the death of massive stars, simply assuming such a return to happen instantaneously when the stellar population is formed. 

 The final term of Eq.~\ref{eq:massevo} characterises the effect of outflow during the star formation process. We adopt the widely used mass loading factor or `wind parameter' to model the strength of the outflow, so that 
\begin{equation}
\label{eq:outflow}
\dot{M}_{\rm out}(t)=\lambda\times\psi(t),
\end{equation}
where the dimensionless quantity `wind parameter' $\lambda$ characterises the strength of the gas removal process relative to the star formation activities. This factor may also evolve over cosmic time, reflecting changes in the physical mechanisms responsible for gas removal, such as stellar/AGN-driven outflows or environmental effects \citep[e.g.,][]{Lian2018mzr,Zhou2022, Zhou2022environment}. However, as a first-order attempt to constrain the underlying physical processes governing the evolution of fast-and slow-rotating galaxies, we assume a time-independent mass loading factor in this work, reflecting the average efficiency of gas removal integrated over the galaxy's lifetime. Exploring potential time-dependence in the mass loading factor will be the subject of future investigation.

Under all the assumptions listed above, we can write down the final equation characterising the evolution of gas mass in a galaxy as:
\begin{equation}
\label{eq:massevo_final}
\dot{M}_{\rm g}(t)=
   A e^{-(t-t_0)/\tau}-S(1-R+\lambda) M_{\rm g}(t).
\end{equation}

In addition to the mass evolution, our model can also describe the evolution of the metal content in the galaxy. To achieve this, in addition to the assumptions mentioned above, we also adopt the instantaneous mixing approximation so that the gas in a galaxy is always  considered well mixed during its evolution. Such assumption allows us to write down the equation of chemical evolution in a galaxy as 
\begin{equation}
\label{eq:cheevo}
\begin{aligned}
\dot{M}_{Z}(t)=& Z_{\rm in}\dot{M}_{\rm in}(t)-Z_{\rm g}(t)(1-R)SM_{\rm g}(t)\\& + y_Z(1-R)SM_{\rm g}(t)
 -Z_{\rm g}(t)\lambda SM_{\rm g}(t).
\end{aligned}
\end{equation}
In the equation, we use $Z_{\rm g}(t)$ to denote the metallicity of the gas cloud, so that $M_{Z}(t)\equiv M_{\rm g}\times Z_{\rm g}$ is the total amount of metals contained in the gas phase.

\begin{figure*}
    \centering
    \includegraphics[width=1.0\textwidth]{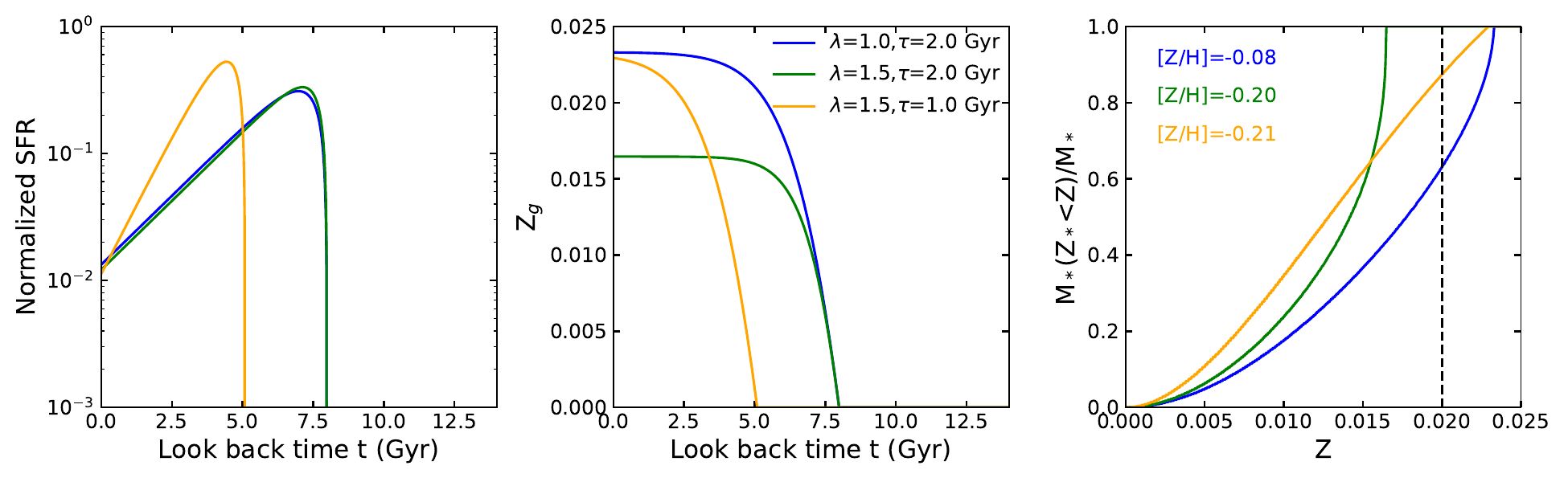}
        \caption{Predicted star formation histories (left), chemical evolution histories (middle), and cumulative metallicity distribution functions (right) for three representative models, with key model parameters indicated in the middle panel. In the right panel, the labels indicate the light-weighted average stellar metallicity for each model.}
     \label{fig:model_prediction}
\end{figure*}

Similar to Eq.~\ref{eq:massevo}, the first term on the right-hand side of this equation represents the gas inflow, with $Z_{\rm in}$ being the metallicity of the infall gas. It is commonly assumed that the infall gas is pristine, so $Z_{\rm in}=0$ is adopted throughout this work. The second term indicates the mass of gas locked up in low-mass stars during the star formation, while the third term represents the chemical-enriched gas returned from dying stars. The enrichment of metals due to the pollution of dying massive stars can be characterised by the so-called metal yield parameter $y_Z$, which is the fraction of metal mass generated per stellar mass. We simply adopt the net yield value $y_Z=0.063$ from \cite{Spitoni2017}, which is calculated through yields of stellar models from \cite{Romano2010} with a Chabrier IMF assumed. Finally, the last term of the equation represents the removal of metal-enriched gas, most likely originating from supernova \citep{Dekel1986} or AGN driven outflows \citep{Silk1998}. To facilitate direct comparison between the stellar and gas-phase metallicities, we adopt the solar oxygen abundance of 8.69 \citep{Asplund2009} throughout this work.

We are fully aware that the model described above represents a simplified description of galaxy evolution. In reality, the trajectories of individual systems in the SFR–stellar mass plane can be highly complex and non-monotonic. However, the aim of the model is not to reconstruct detailed evolutionary paths for individual galaxies, but rather to obtain physically interpretable and self-consistent constraints on the dominant processes that shape the star-formation and chemical-enrichment histories of green-valley systems. When applied statistically, comparisons of the inferred model parameters across different subsamples can provide valuable diagnostic insight, even when the underlying model is simple.

Indeed, similarly simplified approaches have been successfully used in a number of previous studies, including investigations of the mass–metallicity relation in star-forming and passive galaxies \citep{Spitoni2017, Lian2018mzr}, the unusually low metallicities of passive galaxies at $z \sim 0.7$ \citep{Beverage2021}, and the origin of metallicity gradients in local galaxies \citep{Belfiore2019bathtub}. Our earlier analyses using a related framework \citep{Zhou2022environment, Zhou2025} have shown that such models are well suited for tracing broad trends such as quenching timescales, the role of feedback, and the influence of environment. In an on going work (Zhou et al. in prep.),   this framework is also tested using mock galaxies from the Illustris-TNG simulation to systematically examine which semi-analytic prescriptions best recover realistic SFHs and ChEHs and to quantify how parameters are related to physical process that have known "ground-truth". However, as also emphasised in these studies, inferences drawn from models inevitably reflect the assumptions and limitations of the adopted framework, and therefore must be interpreted with appropriate caution.

\subsection{Exploring the parameter space}
\label{ssec:model_prediction}
With the chemical evolution model in place, we are now positioned to explore the parameter space for identifying processes that may have led to the formation of the two types of green valley galaxies with distinct kinematic states. As test cases, we consider three models, all sharing the same present-day SFR but differing in gas accretion histories and outflow strengths.

We begin by constructing a reference model (hereafter referred to as Model 1), which is assumed to have started its gas infall 8 billion years ago ($z\sim1$),
with a gas infall timescale of $\tau=$2 Gyr, and a mild outflow strength characterised by $\lambda=1.0$.
Figure~\ref{fig:model_prediction} presents the predicted evolution of this model galaxy. The first panel displays its SFH (blue line), which indicates an exponential decline in SFR by approximately 1 dex over the past 8 Gyr, thereby bringing the model galaxy into good agreement with the star formation properties observed in present-day green valley galaxies.
In the second and third panels, we display its chemical evolution history and cumulative metallicity distribution function (CMDF), $M_{*}(Z_*<{\rm Z})/M_*$, which represents the mass fraction in stars with a heavy element fraction less than a given value Z. In our simple model, where gas-phase metallicity increases steadily over time, the Z value at which $M_{*}(Z_*<{\rm Z})/M_*$ reaches 1 corresponds to the metallicity of the youngest stellar populations (and the present-day gas from which these stars formed). For comparison, we also compute the light-weighted average stellar metallicity, following the definition adopted by \citet{Sanchez2022pipe3d}, and indicate it on the third panel. These plots reveal that the relatively long timescale of the gas supply, together with the mild outflow strength, allows for sustained chemical enrichment in this galaxy -- from the CMDF we see that nearly 40\% of the stars in
the galaxy have Z>0.02, with an averaged [Z/H]$=-0.08$.

We then construct an alternative model (hereafter Model 2) to represent a galaxy that has experienced a stronger feedback process during its evolution, characterised by a higher mass loading factor of $\lambda = 1.5$. Similar to Model 1, we assume that gas accretion began 8 Gyr ago and adopt the same gas infall timescale of $\tau = 2.0$ Gyr. This configuration reflects a scenario where the enhanced feedback is not sufficiently strong to suppress or delay gas accretion significantly, so Model 1 and Model 2 share the same SFH (see the left panel in Figure~.\ref{fig:model_prediction}). However the higher mass loading factor in Model 2 results in a substantially lower  present-day gas-phase metallicity, along with a systematically lower stellar metallicity distribution and mean stellar metallicity. These results closely resemble the observed trends at the low-mass end of Figure~\ref{fig:pip3d_metal}, where galaxies exhibit reduced metal enrichment. 
Comparing these two models underscores the fact that, despite sharing an identical gas infall history (and thus SFH), variations in outflow strength alone can lead to pronounced differences in the chemical composition of galaxies.

Additionally, we introduce a third model (Model 3) to explore an alternative evolutionary scenario. In this case, the galaxy is assumed to have undergone a similarly strong feedback process ($\lambda = 1.5$) as in Model 2, but coupled with a more intense and short-lived episode of star formation. Unlike Model 2, the feedback in Model 3 is assumed to be sufficiently strong to influence the gas accretion process itself. We adopt a shorter infall timescale of $\tau = 1.0$ Gyr, and to achieve the same present-day star formation rate as the reference model (Model 1), we delay the onset of gas infall to a lookback time of 5.1 Gyr. This combination yields a star formation history that declines more steeply, while still matching both the current SFR and gas-phase metallicity of the reference model. The predicted SFH, ChEH, and CMDF of Model 3 are shown as the orange lines in Figure~\ref{fig:model_prediction}.

Interestingly, the CMDF of Model 3 (orange line in the third panel of Figure~\ref{fig:model_prediction}) shows that, despite reaching a present-day gas-phase metallicity comparable to that of the reference model, it forms significantly fewer metal-rich stars over its evolutionary history—only $\lesssim 20\%$ of the stellar mass is in stars with $Z > 0.02$, with an averaged [Z/H]$=-0.21$. This outcome can be easily understood within the context of the model: the short gas infall timescale limits dilution from pristine inflowing gas, allowing the gas-phase metallicity to rapidly reach values similar to those in model 1. However, the stronger outflow efficiently expels a significant fraction of the newly synthesised metals, thereby suppressing the overall chemical enrichment of the stellar component. As a result, although the present-day gas-phase metallicity appears similar between the two models, their stellar metallicity distributions differ substantially. This scenario reproduces the behaviour observed at the high-mass end of Figure~\ref{fig:pip3d_metal}, highlighting that comparable present-day gas-phase metallicities can be reached through distinct evolutionary pathways. These different pathways leave clear signatures in the stellar populations, shaped by the combined effects of gas inflow timescales and outflow efficiencies.

In summary, our simple model possess sufficient flexibility to reproduce the observed variations of metallicity
in both gas and stellar phases between the faster and slower populations, as shown in Figure \ref{fig:pip3d_metal}. However, it remains to be confirmed whether this simplified scenario can successfully 
produce galaxy spectra consistent with those observed in the MaNGA survey. Fortunately, our semi-analytic fitting approach allows us to directly fit the chemical evolution model to the observed spectra of individual galaxies, enabling us to assess the validity of these evolutionary scenarios on a galaxy-by-galaxy basis. From the resulting best-fit models, we can extract key information about their gas infall and outflow histories, providing valuable insights into the physical processes driving galaxy evolution.

\subsection{Fitting to the stacked spectra}
\label{subsec:spectral_fitting}

\begin{figure*}
    \centering
    \includegraphics[width=1.0\textwidth]{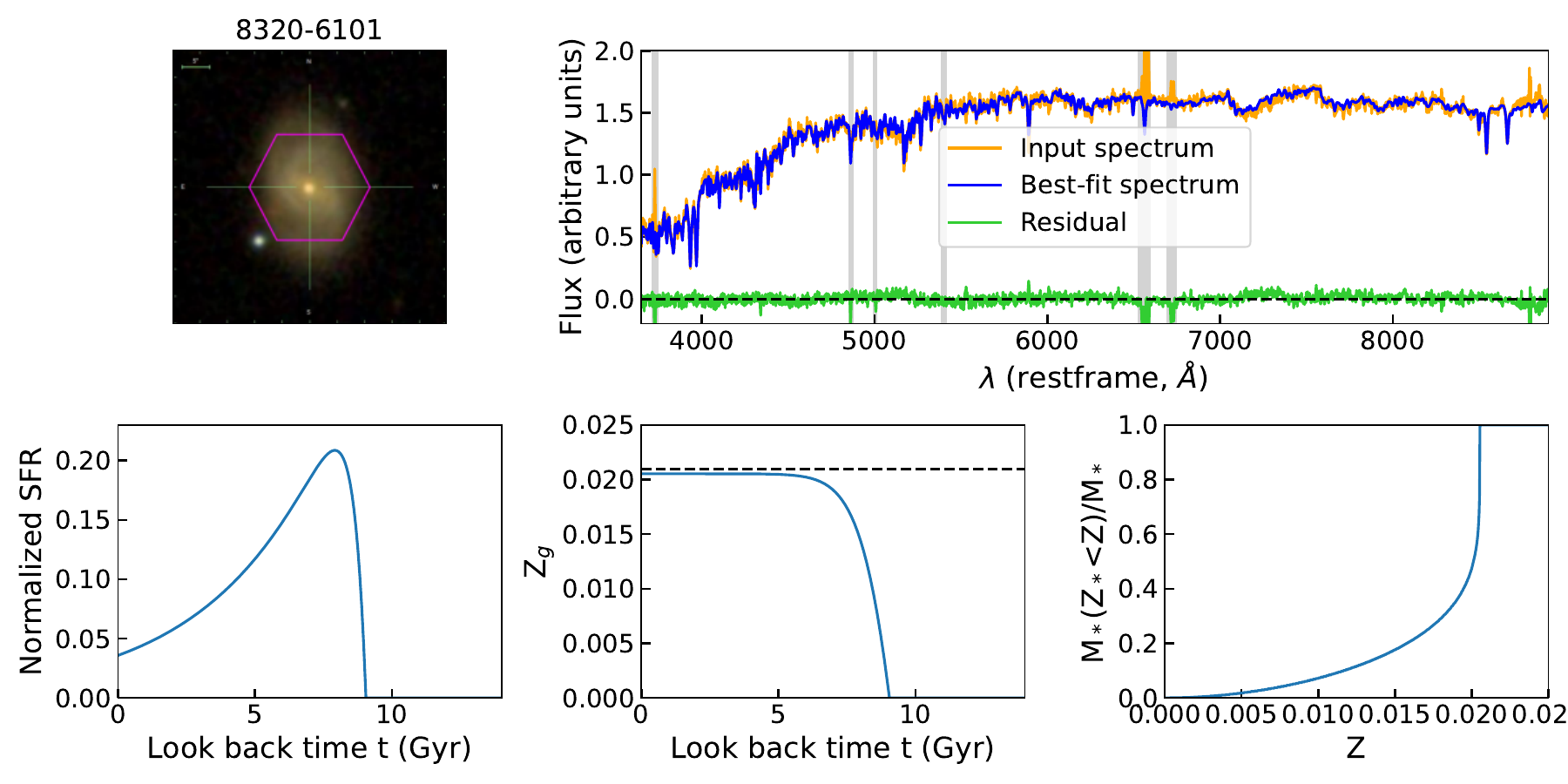}
        \caption{Example of the spectral fitting process to a green valley galaxy in our sample. The top-left panel shows the optical image of the galaxy, with its MaNGA plateifu ID indicated. The top-right panel shows the best-fit model to the observed data. In this panel, the orange line is the observed spectrum stacked within 1 Re of the galaxy, while the blue line shows the best-fit model spectrum. At the bottom, a green line indicates the residuals, with the grey shaded region showing the emission lines that are masked during the fitting process. The bottom panels show the SFH (left), ChEH (middle) and CMDF (right) of this galaxy as calculated from the best-fit parameters. In the middle panel at the bottom, the grey dashed line indicates the observed gas-phase metallicity as obtained from emission line analysis.}
     \label{fig:exmaplefitting}
\end{figure*}

With our general model for predicting a galaxy’s SFH and ChEH established, we now proceed to constrain its parameters using galaxy spectral data.
We perform this fitting process under a Bayesian framework, which has been presented and tested in detail in \cite{Zhou2022}. In short, we first generate a set of model parameters, including those related to the chemical evolution model and an additional dust attenuation parameter required in the stellar population synthesis procedure, from an appropriate prior distribution (listed in Table \ref{tab:paras}). Parameters for the chemical evolution model are used to calculate the SFH and ChEH following Eq.~\ref{eq:massevo} and Eq.~\ref{eq:cheevo}, during which the present-day gas phase metallicity $Z_{\rm g}$ is obtained from the ChEH.  We calculate a model spectrum corresponding to the model SFH and ChEH using the standard stellar population synthesis approach (see review by \citealt{Conroy2013}) and the same E-MILES SSP models as used in the initial {\tt pPXF} analysis (see section ~\ref{ssec:global}). During the calculation, we use a simple screen dust model specified by a \cite{Calzetti2000} attenuation curve.

Before fitting the observed spectrum with the model templates, it is essential to account for the broadening of the observations caused by both stellar velocity dispersion and instrumental effects. This is achieved by applying the necessary broadening to the E-MILES templates using the velocity dispersion derived from the {\tt pPXF} fit. In addition, we mask strong emission lines identified during the {\tt pPXF} fitting process. After these preprocessing steps, we compare the model spectrum and the predicted current gas-phase metallicity to the MaNGA data using a $\chi^2$-like likelihood function:

\begin{equation}
\label{likelyhood}
\ln {L(\theta)}\propto-\sum_{i}^N\frac{\left(f_{\theta,i}-f_{\rm D,i}\right)^2}{2f_{\rm err,i}^2}-
\frac{(Z_{\rm g,\theta}-Z_{\rm g,D})^2}{2\sigma_{Z}^2
},
\end{equation}
where $f_{\theta, i}$ and $f_{\rm D, i}$ are the flux predicted from the model with parameter set $\theta$ and from the observed spectrum respectively, with $f_{\rm err,i}$ being the corresponding error spectrum. The sum is made over all $N$ wavelength points. Similarly, $Z_{\rm g,\theta}$ and $Z_{\rm g,D}$ are the gas phase metallicity from the model and observed data, with $\sigma_{Z}$ being the uncertainty in the gas phase metallicity estimates,  which was set to $\sigma_{Z}=0.1Z_{g,D}$ \citep{Zhou2022}. We use the {\tt MULTINEST} sampler \citep{Feroz2009, Feroz2013} and its \textsc{Python} interface \citep{Buchner2014} to explore the parameter space and the posterior distributions. We obtain the best-fit model parameters, which are used 
to calculate the corresponding SFH, ChEH and CMDF. In Figure \ref{fig:exmaplefitting}, we present the result obtained when fitting our models to the observed spectrum for an example galaxy with MaNGA ID plateifu 8320-6101. The optical image reveals that this galaxy appears relatively red in colour, yet still shows spiral structure, indicating that it is a typical green valley galaxy undergoing a transition to the red sequence. Our model fits both the observed spectrum and the current gas-phase metallicity of the galaxy reasonably well, providing confidence that the model has captured the key physical processes driving its evolution. 
Comparable fit quality is achieved across the full faster and slower populations, enabling a statistically robust comparison of their evolutionary pathways. In addition, our tests reveal that adopting a different SSP template, such as BC03 \citep{BC03}, only leads to systematic shifts in parameters like the infall starting time and outflow strength, but the overall trends with stellar mass and kinematic state remain unchanged. In the following sections, we analyse these results to identify the physical mechanisms that have shaped the formation and evolution of green valley galaxies.

\begin{table}
	\centering
	\caption{Priors of model parameters used to fit galaxy spectra}
	\label{tab:paras}
	\begin{tabular}{lccr}
		\hline
		Parameter & Description & Prior range\\
		\hline
		$\tau$ & Gas infall timescale & $[0.0, 14.0]$ Gyr\\
		$t_{0}$ & Start time of gas infall & $[0.0, 14.0]$ Gyr\\
		$\lambda$ & The wind parameter & $[0.0, 5.0]$\\
		$E(B-V)$&  Dust attenuation parameter & $[0.0, 0.5]$\\
		\hline
	\end{tabular}
\end{table}

\section{Results}
\label{sec:results}

\begin{figure*}
    \centering
    \includegraphics[width=0.9\textwidth]{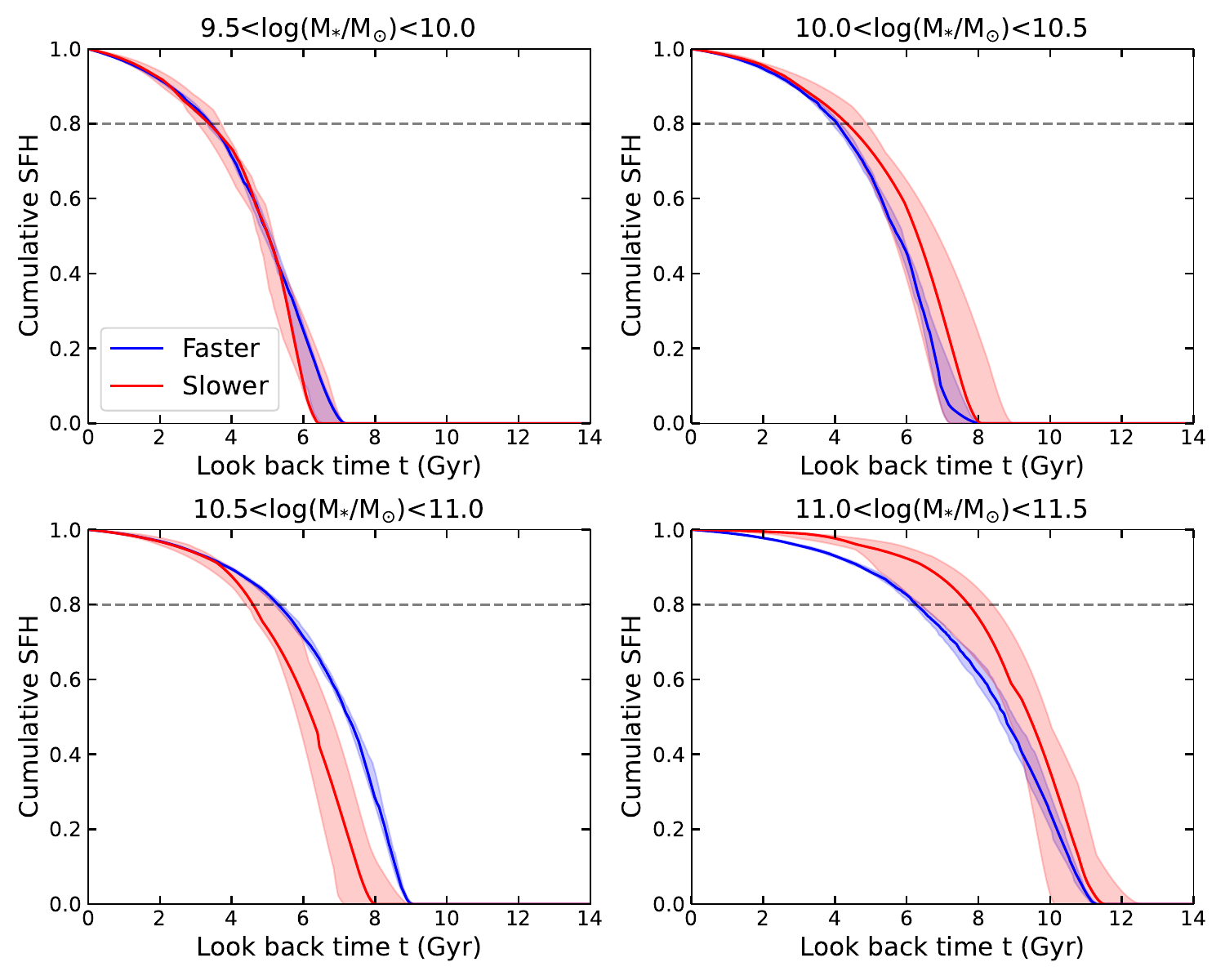}
        \caption{The average cumulative SFH of our sample galaxies.
        Different panels show results for different stellar mass bins, as indicated. In each bin, the blue line shows the results for the faster population, while the red lines are for the slower population. Shaded regions around lines indicate the uncertainty obtained from the standard deviation of 1,000 bootstrap resamplings.}
     \label{fig:result_SFH}
\end{figure*}

\subsection{The mass growth history}
We show in Figure~\ref{fig:result_SFH} the average cumulative SFHs of the faster and slower populations, derived from model fits to the observed spectra and gas-phase metallicities, for each of the four stellar-mass bins considered. The plot shows a systematic trend with stellar mass: as stellar mass increases, both the two populations begin their star formation earlier and accumulate stellar mass over shorter timescales, as indicated by the steeper rise in mass over time. This is consistent with the general phenomenon of "downsizing" in galaxy formation, which has been extensively investigated in previous studies \citep{Panter2003, Panter2007, Kauffmann2003, Heavens2004, Fontanot2009, Muzzin2013}. Our results suggest that galaxies, regardless of their kinematic state, broadly follow this downsizing formation trend.

Turning to the differences between the faster and slower populations, it is interesting to note the contrasting behaviours observed at low and high stellar mass bins, as shown in Figure \ref{fig:result_SFH}. In the two lowest stellar mass bins ($10^{9.5}<M_*/{\rm M}_{\odot}<10^{10.5}$), the two populations exhibit almost identical SFHs. In contrast, 
a difference in the cumulative SFHs between the faster and slower populations emerges in higher stellar mass bins. In the most massive bin ($10^{11.0}<M_*/{\rm M}_{\odot}<10^{11.5}$)
, the best-fit model yields relatively short star-formation timescales for the slower population, such that they accumulate approximately 80\% of their stellar mass within the first 3 billion years of their star formation. In contrast, the faster population form their stars over a much longer timescale, spending approximately 5 billion years to accumulate 80\% of their stellar mass.

\begin{figure}
    \centering
    \includegraphics[width=0.45\textwidth]{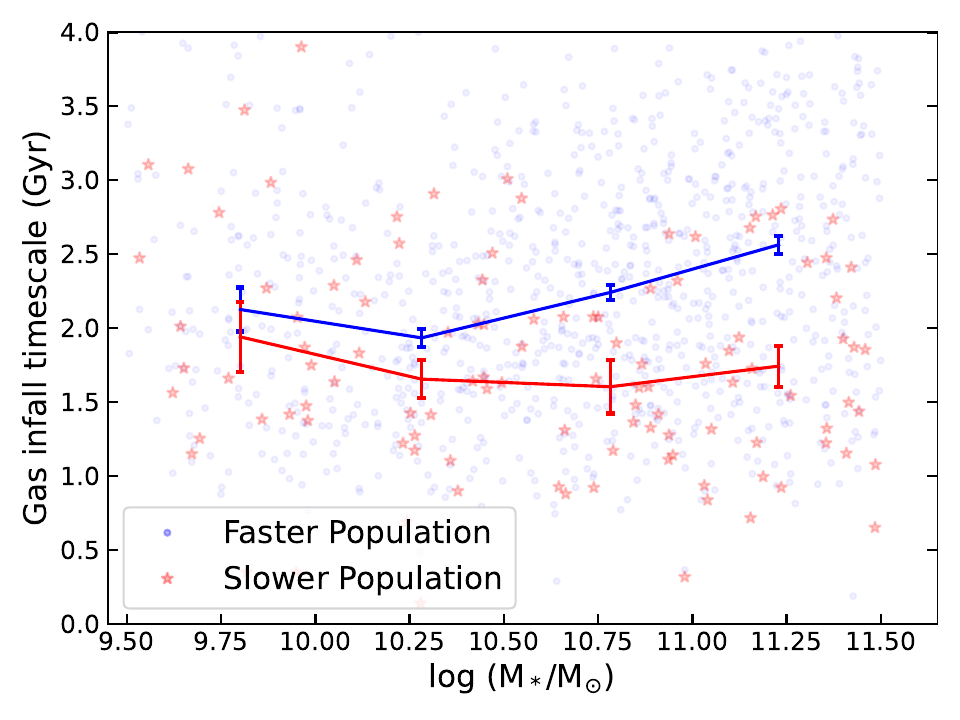}
        \caption{The gas infall timescale obtained from the best-fit models. Fast-rotating galaxies are shown as blue dots while slow-rotating galaxies are shown in red. The colour lines indicate the corresponding mean average values in four stellar mass bins, with uncertainty obtained from the standard deviation of 1,000 bootstrap resamplings.}
     \label{fig:result_gasinfall}
\end{figure}

To quantify these differences and aid in their interpretation, we utilise the best-fit parameters derived from our chemical evolution models. In this framework, the parameter most directly linked to the SFH of a galaxy is the gas infall timescale, $\tau$. In Figure~\ref{fig:result_gasinfall}, we present the best-fit $\tau$ values for our sample galaxies as a function of their stellar mass. For low-mass galaxies ($10^{9.5}<M_*/{\rm M}_{\odot}<10^{10.5}$), both the faster and slower populations exhibit comparable gas infall timescales, with typical values around $\tau \sim 2$ Gyr, indicating little difference between the two kinematic classes at this mass scale. However, at higher stellar masses, a systematic divergence emerges: the slower population tend to have significantly shorter infall timescales than the faster population of similar mass. Specifically, among the most massive galaxies in our sample ($10^{11.0}<M_*/{\rm M}_{\odot}<10^{11.5}$), fast-rotating galaxies exhibit an average $\tau \sim 3$ Gyr, while slow-rotating galaxies show a shorter average timescale of $\tau \sim 1.8$ Gyr. These trends are entirely consistent with the star formation histories inferred in Figure~\ref{fig:result_SFH}, further supporting the notion of distinct gas accretion and star formation pathways for the faster and slower populations at the massive end.

Interestingly, the trends derived from our best-fit models align well with the initial speculation presented in Section~\ref{ssec:sample} in the context of Table~\ref{tab:bins}. There, we proposed that the fraction of galaxies observed in the green valley is linked to the speed at which they transition from star-forming to passive, potentially explaining the differing green valley fractions between the faster and slower populations. The inferred SFHs and gas infall timescales now provide support for this scenario: massive slow-rotating galaxies on average end their star formation activity more rapidly than their fast-rotating counterparts, naturally leading to a lower fraction of green valley slow-rotating galaxies relative to the total population. At the low-mass end, the differences in SFHs and gas infall timescales between the faster and slower populations are less pronounced, consistent with the more modest disparity in their green valley fractions observed in Table~\ref{tab:bins}.

\begin{figure*}
    \centering
    \includegraphics[width=0.95\textwidth]{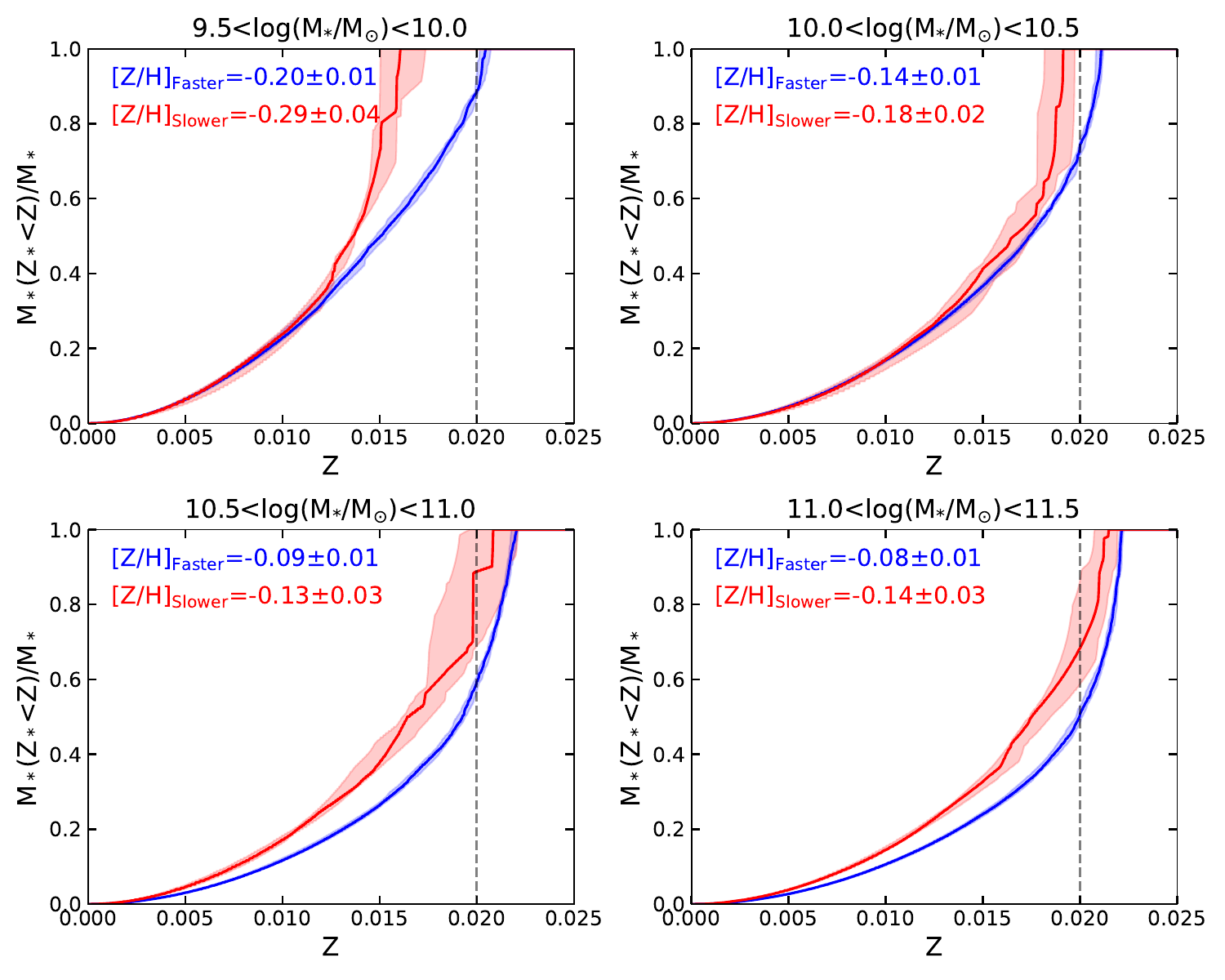}
        \caption{The average CMDF of our sample galaxies. Different panels show results for different stellar mass bins, as indicated. In each bin, the blue line shows the results for the faster population, while the red lines are for the slower population. Shaded regions around lines indicate the uncertainty obtained from the standard deviation of 1,000 bootstrap resamplings, while the labels indicate the light-weighted average stellar metallicity for the sample considered.}
     \label{fig:result_CMDF}
\end{figure*}

\subsection{The chemical composition}
We now turn to the chemical composition of our sample galaxies as probed by the best-fit models. We begin by examining the metal content in the gas phase, which, by construction in our modelling framework, corresponds to the metallicity of the youngest, most metal-rich stellar populations. In our best-fit models, this value can be inferred from the point where M$_*$(Z$_*$<Z)/M$_*$ reaches 1 on the CMDF. We note that the observed present-day gas-phase metallicity serves as a constraint during the fitting process; however, this does not imply that the model outputs will exactly reproduce the input gas-phase metallicities. Instead, the fitting algorithm seeks a compromise between the constraints from the current gas-phase metallicity and the stellar spectra, resulting in a best-fit gas-phase metallicity with a median deviation of approximately 20\% from the input values. As shown in Figure~\ref{fig:result_CMDF}, at the low-mass end, the slower population tends to exhibit systematically lower gas-phase metallicities than the faster population. In contrast, at the high-mass end, the difference becomes negligible. This trend is broadly consistent with that observed in the input gas-phase metallicities shown in Figure~\ref{fig:pip3d_metal}, suggesting that the fitting procedure achieves a reasonable balance between the stellar and gas-phase constraints and retains the overall trend present in the data.

Beyond the chemical content of the gas phase, an even more intriguing phenomenon emerges in the distribution of metals within stars, as revealed by the overall shape of the best-fit CMDF. For the least massive galaxies ($10^{9.5} < M_*/{\rm M}_{\odot} < 10^{10.0}$), although the faster and slower populations exhibit nearly identical SFHs as shown in Figure \ref{fig:result_SFH}, they follow different evolutionary paths in their chemical enrichment processes. While their CMDFs are similar at low metallicities, they begin to differ markedly at the metal-rich end, ultimately producing distinct stellar metallicity distributions and present-day gas-phase metallicities. This distinction between the faster and slower populations is well reproduced by the model predictions of Model 1 and Model 2, illustrated by the blue and green lines in Figure \ref{fig:model_prediction}. The models indicate a stronger outflow in the slower population relative to the faster population, while the gas infall timescales remain unchanged. 

Moving to the high-mass end, we observe a different behaviour in how the CMDFs vary between the faster and slower populations. For the most massive galaxies ($10^{11.0}<M_*/{\rm M}_{\odot}<10^{11.5}$), although the two populations exhibit nearly identical present-day gas-phase metallicities, their stellar metallicity distributions differ markedly. Fast-rotating galaxies host a significantly higher fraction of metal-rich stars, with approximately 60\% of their stellar mass residing in stars with $Z > 0.02$, compared to only $\sim$40\% in the slower population. This leads to a higher average stellar metallicity in the faster population, consistent with the trends seen in Figure~\ref{fig:pip3d_metal}, despite their similar gas-phase metallicities. When considered alongside the SFHs shown in Figure~\ref{fig:result_SFH}, these differences are well reproduced by the model predictions of Model 1 and Model 3, represented by the blue and orange lines in Figure~\ref{fig:model_prediction}. In this scenario, massive slow-rotating galaxies experienced not only stronger gas removal but also shorter gas infall compared to their fast-rotating counterparts. Their SFHs indicate intense, short-lived episodes of star formation early in their evolution, followed by a sharp decline in gas supply.

\begin{figure}
    \centering
    \includegraphics[width=0.45\textwidth]{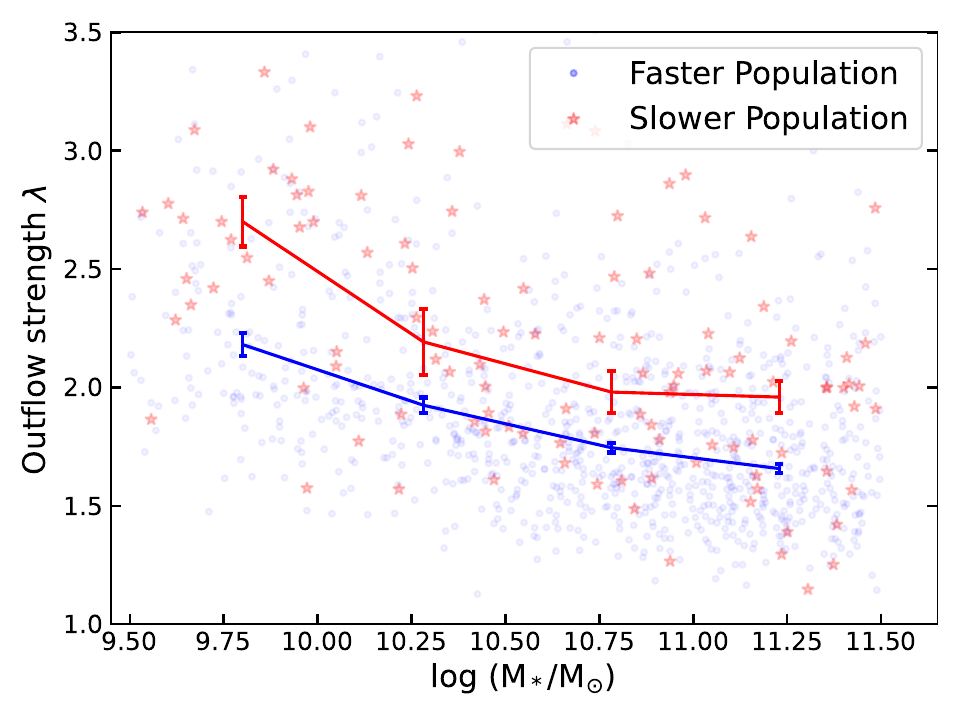}
        \caption{The outflow strength characterising using the mass-loading factor $\lambda$, obtained from the best-fit models. Fast-rotating galaxies are shown as blue dots while slow-rotating galaxies are shown in red. The colour lines indicates the corresponding mean average values in four stellar mass bins, with uncertainty obtained from the standard deviation of 1,000 bootstrap resamplings.}
     \label{fig:result_outflow}
\end{figure}

 To quantitatively illustrate the variation in outflow strength across our sample, we present the mass-loading factor $\lambda$, obtained from the best-fit models, in Figure \ref{fig:result_outflow}. The mass-loading factor has been widely used in the literature to probe key aspects of galaxy evolution—for example, to investigate the stronger galactic winds thought to operate in local star-forming galaxies compared to passive systems \citep{Spitoni2017}, to explain the stellar and gas-phase mass–metallicity relations \citep{Lian2018}, and to reproduce the observed G-dwarf metallicity distributions in star-forming galaxies \citep{Spitoni2021Gdwarf}. These studies demonstrate that the mass-loading factor is a useful and physically motivated indicator of the efficiency of gas removal processes in galaxies.
From our model-based analysis on these green valley galaxies, we first observe in Figure \ref{fig:result_outflow} a significant mass dependence of the outflow strength parameter in both the faster and slower populations. The least massive galaxies have $\lambda\sim2.5$, while for the most massive galaxies, this value drops below 2. 

We note here that, despite differences in the absolute, model-dependent values of the 
mass-loading factors, theoretical studies of stellar feedback such as the 
analytic calculations in \citet{Hayward2017} and the FIRE simulation 
\citep{Muratov2015}, consistently predict an anti-correlation between galaxy mass and mass-loading factor that closely resembles the trend we derive empirically from our model. This agreement across independent approaches increases our confidence that, although our framework is simplified, the fitted mass-loading factors are likely capturing a physical signal linked to gas-removal processes in galaxies. We return to this point in the discussion section.

In addition to this global trend, the model suggests that the slower population, on average, experience stronger outflows compared to the faster population. This trend is present across the entire mass range covered by our sample. These results are entirely consistent with expectations based on the CMDFs, where enhanced outflows in the slower population play a key role in shaping their chemical evolution.

\subsection{Comparison with empirical evidence}

These results derived from our model fitting portray an interesting picture. However, model-based inferences often suffer from limitations and degeneracies inherent to the adopted framework. We therefore seek additional empirical evidence to support this scenario. The model based analysis indicates that the faster and slower populations exhibit different evolutionary timescales at the high-mass end. Empirically, the abundance of $\alpha$-elements, which are found to be higher in galaxies formed over shorter timescales of formation \citep[e.g.,][]{Worthey1994, Thomas2005, Zheng_etal2019}, is commonly used as a proxy for the timescale of galaxy formation. We therefore measure the Lick indices Mgb, Fe5270, and Fe5335 from both the observed spectra and the best-fit continua obtained from the {\tt pPXF} fits presented in Section~\ref{ssec:global}.

From these measurements, we computed the index ratio Mgb/$\langle$Fe$\rangle$ $\equiv 2 \times$ Mgb / (Fe5270 + Fe5335). Under the assumption that all $\alpha$-elements can be tracked by Mg, such a ratio serves as a well-established proxy for the relative abundance of $\alpha$-elements. 
Since the SSP models used in the {\tt pPXF} analysis assume solar abundance ratios, we adopt as a proxy of [$\alpha$/Fe] the ratio of the measured Mgb/<Fe> with respect to the same quantity for best-fit spectra  provided by {\tt pPXF}. This approach ensure that the
$\alpha$-enhancement estimate does not rely on the analytic chemical evolution model and therefore serves as an independent consistency check.

\begin{figure}
    \centering
    \includegraphics[width=0.45\textwidth]{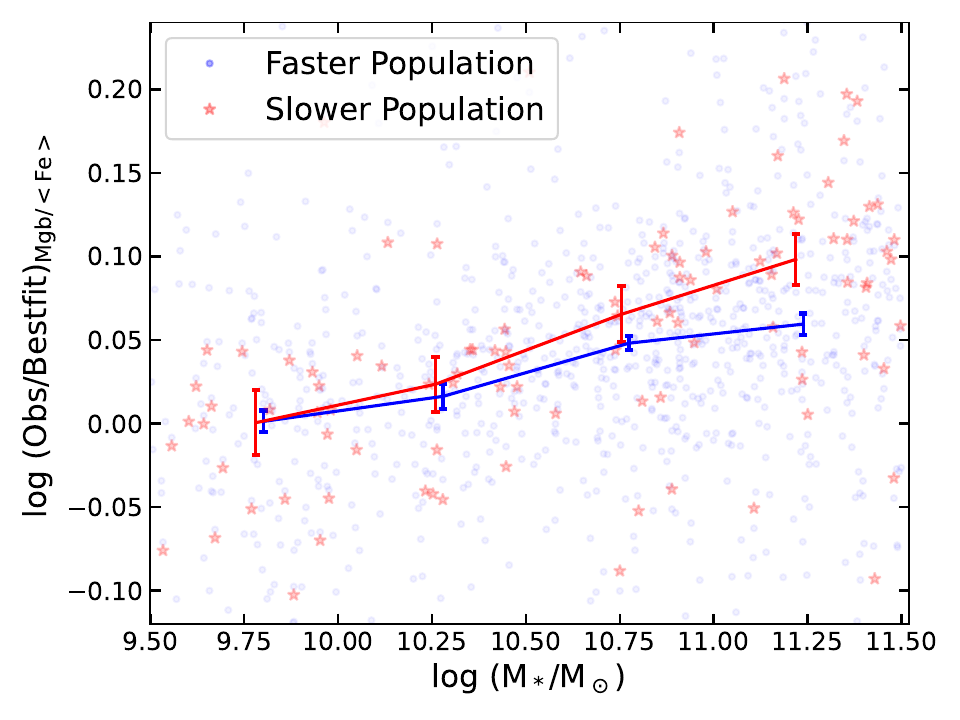}
        \caption{The ratio between Mgb/<Fe> measured from the observed spectra and best-fit model spectra. Fast-rotating galaxies are shown as blue dots while slow-rotating galaxies are shown in red. The colour lines indicate the corresponding mean average values in four stellar mass bins, with uncertainty obtained from the standard deviation of 1,000 bootstrap resamplings.}
     \label{fig:result_mgbfe}
\end{figure}

In Fig.~\ref{fig:result_mgbfe}, we plot this ratio as a function of stellar mass for our sample galaxies. We find that at lower stellar masses, both the two populations exhibit nearly solar $\alpha$-element abundances, while at the high-mass end, the slower population show systematically higher $\alpha$-enhancement compared to the faster population of similar stellar masses. In fact, similar tendency for slow rotators to exhibit higher $\alpha$-enhancement than fast rotators has 
been reported previously, for example in the ATLAS$^{\mathrm{3D}}$ survey 
\citep{McDermid2015} and in MaNGA \citep{Bernardi2019}. Although based on slightly different sample definitions, our results are broadly consistent with these findings and further reveal a clear mass dependence in the differences when examined green-valley galaxies in different stellar mass bins.
This empirical evidence easily fits with the results of our spectral modelling results, suggesting that, especially in the highest mass range, the star formation timescales correlates with kinematic properties.

By synthesising the results from our best-fit models as well as empirical evidence, we obtain a more coherent picture of how fast- and slow-rotating galaxies evolve across the green valley. At the low-mass end, model results indicate comparable gas infall timescales, resulting in similar past SFHs for both populations. At the same time, model results suggest that the slower population tend to experience stronger outflows, which would remove a larger fraction of metals from the system and may lead to lower metallicities in both the stellar and gas phases compared to the faster population. In contrast, at the high-mass end, the model suggests that the evolutionary timescales diverge as a function of kinematic morphology-- the slower population exhibit shorter gas infall timescales and stronger outflows relative to the faster population of similar stellar mass. 
In the model framework,
this combination results in a more rapid quenching process, consistent with the lower observed fraction of the slower population in the green valley. 
In addition, the balance between the
reduced dilution from pristine infall due to the short-lived gas supply and the efficient removal of chemically enriched material allows slow-rotating galaxies to reach gas-phase metallicities comparable to those of the fast-rotating galaxies at the present day, while still leaving behind more metal-poor stellar populations, as also revealed by the non-parametric spectral fitting results in Fig~\ref{fig:pip3d_metal}. This combined analysis of stellar and gas-phase metallicities, supported by a tailored chemical evolution model and robust spectral fitting, provides a powerful diagnostic of divergent evolutionary pathways. These findings offer important insight into the underlying physical mechanisms governing the distinct formation histories of the faster and slower populations, which we will explore in more detail in the following section.

\section{Discussion}
\label{sec:discussion}
\subsection{Kinematic structure as a tracer of galaxy evolution}
The distinct populations of green valley galaxies and the variations in their quenching paths have been explored in numerous previous studies, using methods ranging from colour-colour diagram analysis (e.g., \citealt{Schawinski2014}) to more detailed investigations of stellar population properties (e.g., \citealt{Peng2015}, \citealt{Carnall2018}, \citealt{Wang2024Nat}).
In the work of \cite{Schawinski2014}, green valley galaxies are divided into early and late types based on morphology classifications from Galaxy Zoo \citep{Lintott2011}. By investigating the different distributions of early- and late-type green valley galaxies on dust-corrected UV–optical colour–colour diagrams, they reveal variations in the quenching timescales of the two types. From these results, they propose that late-type galaxies quench through the slow consumption of their gas reservoirs. In contrast, early-type galaxies may have undergone major mergers that disrupt their discs, leading to violent starbursts that quickly consume their gas and quench star formation in relatively short timescales. A similar narrative is proposed in \cite{Wang2024Nat}, where morphology-based classifications are replaced by a kinetic-based classification. Thanks to the IFU data from MaNGA, galaxies are divided into two populations according to their angular momentum. By analysing recent star formation activity and stellar metallicities, the authors conclude that fast-rotating galaxies are primarily quenched through the long-timescale consumption of their gas reservoirs, while slow-rotating galaxies likely quench via rapid gas removal processes.

The morphology of a galaxy is a strong indicator of its past evolution, as disk galaxies and ellipticals are found to form through different pathways \citep{Mo1998}. However, when a galaxy undergoes quenching, the fading of its disk component makes it harder to distinguish between early- and late-type galaxies. In addition, studies have revealed the presence of pseudo-bulges in galaxies \citep{Carollo1997}, further complicating the classical bulge–disk distinction. In some cases, components identified as “bulges” may in fact exhibit disk-like structural or kinematic properties. The introduction of kinematic information proves to be a powerful tool for clearly disentangling different kinds of bulge systems \citep{Hu2024}. Additionally, \cite{Wang2024Nat} demonstrate that even in fully quenched systems, one can still differentiate between rotation-supported and velocity-dispersion-supported systems using kinematic data. By adopting kinematic information as the primary criterion for classifying galaxies into fast- and slow-rotating galaxies, we aim to more clearly and robustly uncover differences in their past evolutionary histories. Nevertheless, as a sanity check, we also examine the morphological properties of our samples using the MaNGA Morphology Deep Learning DR17 Catalogue \citep{Sanchez2022}. We find that the faster population are predominantly late-type, although a non-negligible fraction exhibit T-type < 0, indicative of early-type morphology. This consistency underscores the close interplay between galaxy morphology and kinematic state.

The angular momentum of galaxies originates from the material they accrete. In modern theoretical and simulation frameworks, the evolution of individual systems can be complex and non-monotonic, reflecting a broad range of assembly histories and environmental influences. Nonetheless, a consistent picture emerges from cosmological and idealized simulations. Although mergers do not always lead to angular-momentum loss since gas-rich or corotating mergers can transfer substantial orbital angular momentum to the remnant and even increase its stellar specific angular momentum, they remain an efficient pathway for producing low-angular-momentum systems through effective angular-momentum redistribution \citep{Naab2014,Penoyre2017,Lagos2018}. Accordingly, numerical simulations such as \citet{Naab2014} suggest that fast rotators may form from disk-dominated, gas-rich progenitors that avoid disruptive major mergers or subsequently rebuild angular momentum through later gas accretion, whereas slow rotators may be more likely to arise from mergers that efficiently redistribute angular momentum and suppress late-time gas inflow. These distinct evolutionary paths are expected to leave measurable imprints in stellar and gas-phase properties. By fitting to the spectral data, our model-based approaches are capable of providing general, although simplified, histories of the mass growth, chemical enrichment, and gas inflow/outflow of real galaxies. The different results of our model when applied to green-valley galaxies with similar stellar mass (and SFR) but different kinematics, may thus provide a possible link between observed properties and the underlying physical mechanisms predicted by simulations.

\subsection{Distinct evolutionary pathways of the faster and slower populations from our model-based analysis}

Our model fits to the observed spectra suggest a broadly consistent, yet more nuanced, picture compared to earlier studies such as \citet{Schawinski2014} and \citet{Wang2024Nat}. Our results for the slower population indicate that, consistent with their lower metallicities, these systems exhibit systematically higher mass-loading factors across the entire stellar-mass range when compared to the faster population. Within our modelling framework, this implies that the slower population have experienced more efficient gas-removal processes. Combined with the merger-driven origin of the slower population suggested by simulations, this raises the possibility that merger-triggered mechanisms, such as enhanced AGN activity or merger-induced starbursts \citep{Debuhr2012, Garcia-Burillo2015}, may contribute to the inferred differences. However, a more complex picture emerges when considering the full interplay between gas inflow, star-formation history, and stellar mass, especially given the clear mass dependence observed in these properties.

At the high-mass end, \citet{Wang2024Nat} showed that the faster population become progressively more metal-rich when selected from the star-forming sequence, the green valley, and the quenched population, whereas the slower population display nearly constant stellar metallicities regardless of their star-formation state. Our model results offer a natural explanation for this behaviour: from the  shortened gas infall timescales and rapid quenching of star formation activity, the model suggests that these galaxies form their first generation of stars rapidly, and their gas reservoirs are quickly enriched to gas-phase metallicities similar to those of the faster population. However, the rapid exhaustion or suppression of gas supply after the star-formation peak prevents the formation of later generations of more metal-rich stars, leading to nearly stagnant stellar metallicities as they evolve. In contrast, massive fast-rotating galaxies quench more gradually as suggested by the SFHs derived from our best-fit models. The continued gas inflow and extended star-formation activity allow these galaxies to form additional generations of stars from increasingly enriched gas, producing the steadily rising stellar metallicities reported by \citet{Wang2024Nat}. Our semi-analytic spectral fitting, calibrated using high-quality integrated spectra, thus provides a physically motivated interpretation that links the observed behaviours of both stellar and gas-phase metallicities to the distinct evolutionary pathways of massive fast- and slow-rotating galaxies.

The picture diverges at the low-mass end. The best-fit models implies that, although 
low-mass slow-rotating galaxies exhibit stronger outflows than their fast-rotating counterparts, they have similarly long gas infall timescales and comparably extended star formation histories.  In this regime, the elevated outflow strength alone results in lower metallicities in both the stellar and gas phases of slow-rotating galaxies. Nevertheless, the similarly prolonged star formation suggests that their stellar metallicities should steadily increase as they evolve through the green valley, mirroring the trend observed in the faster population. Intriguingly, this is supported by Figure 4 of \cite{Wang2024Nat}, which shows that the lowest-mass star-forming slow-rotating galaxies have significantly lower stellar metallicities than their green valley and passive counterparts.

Combining the consistent evidence from the fraction of green-valley galaxies, our model-derived evolutionary histories, and the $\alpha$-element abundance, we find that the star-formation or gas-supply histories of fast- and slow-rotating galaxies differ significantly only at the high-mass end. At lower masses, despite the presence of a metallicity deficit in the slower population, the two kinematic classes show negligible differences in the timescales of star formation. Given that simulations suggest the key distinction between the origins of the faster and slower populations lies in their assembly histories, this mass-dependent reversal hints that different physical processes may be triggered or become dominant during mergers in low- versus high-mass galaxies, even if both pathways can ultimately lead to the same kinematic transformation.

A natural interpretation is that different feedback processes associated with mergers play a key role. It has been demonstrated that stellar feedback alone cannot fully deplete a galaxy’s gas reservoir, and that only with the inclusion of AGN feedback—capable of completely removing gas—can rapid and sustained quenching be achieved \citep{Springel2005,Schawinski2009,Dubois2013,Prieto2017}. Observations have established a tight correlation between the mass of a galaxy’s central supermassive black hole and the stellar mass of its bulge \citep{Haring2004}. In this case, at the low-mass end, where galaxies possess shallow gravitational potentials and central black holes have yet to undergo significant growth, supernova feedback is likely the dominant mechanism governing their evolution \citep[e.g.][]{Mashchenko2008,Martin-Navarro2018,Dekel2019}. The enhanced outflows observed in our low-mass slow-rotating sample suggest that the proposed merger events responsible for their low angular momentum may also trigger, or be accompanied by, stronger feedback. However, in these systems, the feedback is likely supernova-driven and, as simulations have shown, insufficient to fully expel the gas or suppress ongoing accretion. This is consistent with the similar gas infall timescales and SFHs inferred for the fast and slow populations in our best-fit models. Nevertheless, the imprint of stronger feedback is evident in the lower stellar and gas-phase metallicities observed in the slower population.

In contrast, as the gravitational potential deepens and central black holes become more massive, AGN feedback becomes the dominant mode of regulation above a critical stellar mass, typically around $M_*/{\rm M}_{\odot} \sim 10^{10.5}$\citep[e.g.,][]{Croton2006, Sijacki2007, King2015, Torrey2019, Dekel2019}. For galaxies above this threshold, the stronger outflows observed in the slower population suggest that their low angular momentum nature is likely to originate from a merger event associated with enhanced AGN and star formation activity. This connection is plausible: mergers can generate strong gravitational torques that efficiently drive gas toward the central regions on timescales of a few dynamical times \citep{Hopkins2009,Hopkins2010}. While transporting this gas down to sub-parsec scales to directly fuel black-hole accretion likely requires additional, less well-understood processes, both simulations \citep[e.g.][]{Springel2005} and observations \citep[e.g.][]{Treister2012,Gao2020} provide evidence that mergers might be associated with enhanced AGN and star formation activity that are capable of driving powerful feedback. Such feedback can efficiently evacuate the gas reservoir and suppress further accretion, leading to the rapid quenching of star formation. Consequently, massive slow-rotating galaxies quench significantly faster than their fast-rotating  counterparts, consistent with the divergent SFHs revealed by our best-fit models. Once again, the interplay between gas inflow and outflow leaves clear signatures in the chemical enrichment of both the stellar and gas phases, as reflected in our modelling results.

It should be noted here that, our current model, by design, does not distinguish between different gas-removal mechanisms. Stellar-driven and AGN-driven outflows, as well as external processes such as tidal stripping, can all remove chemically enriched gas from galaxies, lowering their metal content and thereby producing similar effects on the inferred mass-loading factor. While these mechanisms could in principle be modelled separately, accounting for their distinct timescales, energetics, and spatial signatures, the co-added spectra we fit do not provide sufficient constraining power to robustly disentangle them. The simple parametrisation adopted here represents a pragmatic compromise between physical completeness and data limitations. The inferred mass-loading factors should therefore be interpreted as effective, time-averaged gas-removal efficiencies, rather than as direct tracers of any specific feedback mechanism. Nevertheless, the observed dependencies of gas-removal strength on stellar mass and kinematic structure provide important additional clues to the underlying physical processes. When considered alongside trends predicted by theoretical calculations and hydrodynamical simulations, these empirical patterns allow us to make these speculations about the most likely mechanisms shaping the evolution of different galaxy populations. In fact, the possibility of different feedback mechanisms operating at different stellar masses has also been discussed in \cite{Wang2023APJL}. The authors investigated the gas content of the faster and slower populations using HI observations. By examining the mass dependence of HI gas deficiency in the slower population compared to the faster population, they speculated that the observed mass-related variations may indeed be attributed to distinct feedback processes at different mass scale. This provides qualitatively similar scenarios based on entirely independent datasets.

Finally, our findings are broadly consistent with previous studies focusing on quenched early-type galaxies with different kinematic properties \citep[e.g.][]{McDermid2015,Cappellari2016ARA,Smethurst2018,Bernardi2019}. In particular, using similar MaNGA observations, \citet{Bernardi2019} showed that fast rotators are more metal-rich than slow rotators at fixed luminosity and central velocity dispersion. \citet{Smethurst2018} reported that fast rotators quench over a wider range of timescales, whereas quenching in slow rotators is more likely to occur rapidly, consistent with scenarios in which slow rotators form through dynamically fast processes such as major mergers. Although these studies focus on quenched early-type galaxies and adopt the classical fast/slow-rotator classification, the similar trends observed in our green valley sample suggest that such differences in galaxy properties are already established before galaxies become fully quenched, or alternatively that a fraction of our systems may be rejuvenated descendants of quenched populations. Overall, our results place the evolution of kinematically distinct systems within this broader evolutionary framework, and the new approach adopted here provides additional insight into the physical processes driving this evolution.

\subsection{Alternative scenarios}
We acknowledge that alternative mechanisms may contribute to the distinct properties observed between the faster and slower populations in this work. As noted earlier, the mass-loading factor inferred from our model is an effective parameter that could originate from multiple processes causing gas removal from the galaxy, not necessarily internal feedback alone. External mechanisms, most notably ram-pressure stripping \citep{Gunn1972}, can also remove gas from galaxies and thereby mimic the effects of strong outflows. Under such a scenario, the higher mass-loading factors inferred for the slower population could reflect a greater degree of environmentally driven gas stripping rather than more efficient internal feedback.

To examine this possibility, we make use of the environmental information provided in the Galaxy Environment for MaNGA Value Added Catalog (GEMA-VAC; \citealt{Argudo2015}). We explore three environmental indicators for our sample—local density, central/satellite classification, and group richness. None of these quantities show statistically significant or systematic differences between the faster and slower populations.
The literature also presents a diverse picture regarding the environmental dependence of galaxies with different kinematic properties. The ATLAS$^{\mathrm{3D}}$ survey reported an enhanced slow rotator fraction in the densest cluster environments \citep{Cappellari2011ki-morph}, but our MaNGA sample contains few galaxies within such extreme environments. In contrast, results from the MASSIVE Survey indicate that, at fixed stellar mass, the spin parameter shows only weak or no dependence on environment \citep{Veale2017}, consistent with trends seen in cosmological simulations such as EAGLE and HYDRANGEA \citep{Lagos2018}. Although these studies select samples using the classical fast/slow-rotator classification, they highlight a complex picture of the environments of galaxies with different kinematic properties.
Our tentative conclusion, therefore, is that we do not detect a clear environmental difference between the the faster and slower populations in our sample. However, this null result may in part reflect the limited environmental coverage and sample size provided by MaNGA, and we thus expect large future surveys to further shed light on this issue.

Another possible explanation arises from differences in assembly history. Our model-based inference relies on the observed metal deficiency of slow-rotating galaxies relative to fast-rotating ones, yet these two populations are expected to have undergone different assembly pathways, which is not explicitly captured by our modelling. In principle, the slower populations could form through the merger of lower-metallicity progenitors and simply retain those metallicities thereafter, without requiring additional gas outflows. However, in the absence of strong gas removal, ongoing or merger-induced star formation should quickly enrich the ISM \citep{Spitoni2017}. Thus, only mergers that are entirely “dry” with essentially no cold gas and no subsequent star formation could preserve such low metallicities.

While simulations do show that some  galaxies with low angular momentum may form through dry mergers, a substantial fraction are associated with gas-rich mergers \citep{Naab2014, Penoyre2017, Lagos2018}, which would normally trigger starbursts and rapid chemical enrichment unless metals are efficiently expelled. Regarding the origin of the stars themselves, recent results from the Illustris–TNG simulations indicate that in-situ star formation dominates the stellar mass of galaxies up to at least Milky Way mass \citep{Wittig2025}. While the ex-situ fraction increasing at higher masses, in-situ stars are also typically more centrally concentrated \citep{Pulsoni2021, Wittig2025}. As our analysis focuses on stellar populations within 1Re, a region likely more dominated by in-situ stellar mass, the assembly-driven explanation is unlikely to dominate across the full mass range considered here, although it may play a non-negligible role in the highest masses galaxies. Future large-scale simulations with improved mass and spatial resolution should provide further insight into this possibility.

\section{Summary}
In this paper, we investigate the formation and evolution of green valley galaxies using data from the SDSS-IV/MaNGA survey, dividing the sample into the faster and slower populations based on their angular momentum and ellipticity. We measure average stellar and gas-phase metallicities within 1 R$_{\rm e}$ to probe global differences between the two populations. To interpret these differences, we construct a simple yet comprehensive evolutionary model that incorporates star formation, gas infall, and outflows, constrained by stacked spectra and gas-phase metallicities within 1 R$_{\rm e}$. This approach yields self-consistent star-formation and chemical-evolution histories, as well as parameters characterising gas accretion and outflow strengths, providing new insights into
the evolution of green valley fast- and slow-rotating galaxies. Our main conclusions are:

\begin{itemize}
    \item Fast- and slow-rotating galaxies, both selected from the green valley, show differences in their chemical compositions in both the stellar and gas phases within 1 R$_{\rm e}$. In the stellar phase, the slower population are more metal-poor compared to the faster population of similar stellar masses across the entire mass range considered in this work. In the gas phase, however, the slower population are more metal-poor than the faster population only at the low-mass end, while at the high-mass end, the two populations have similar gas-phase metallicities.

    \item Based on a simple but physically motivated model, our semi-analytic spectral fitting approach reveals more detailed differences in the SFHs and chemical compositions of fast- and slow-rotating galaxies. At the low-mass end, the two populations exhibit very similar SFHs, gradually accumulating their stellar masses over long timescales of star formation. However, the chemical enrichment in the slower population is suppressed during their evolution, leading to lower present-day gas-phase metallicities. At the high-mass end, the slower population accumulate their stellar mass and undergo quenching in star formation over a shorter timescale compared to the faster population of similar stellar mass. Although both populations end up with similar gas-phase metallicities at present, the slower population form a smaller fraction of high-metallicity stars than the faster population, showing a metal deficiency in the stellar phase.

    \item The variations in gas infall timescales and outflow strengths from the best-fit models offer additional insights into the divergent evolutionary pathways of fast- and slow-rotating galaxies. Across the entire stellar mass range examined in this study, the slower population consistently exhibit stronger outflows than the faster population. However, gas infall timescales display more nuanced behaviour: while the two populations show similar infall timescales at the low-mass end, massive slow-rotating galaxies exhibit significantly shorter infall timescales compared to fast-rotating galaxies of similar mass. This interplay between enhanced outflows and varying inflow timescales results in the persistent stellar-phase metal deficiency observed in slow-rotating galaxies, as well as the mass-dependent differences in gas-phase metallicity between the two populations.

    \item Within our modelling framework, these results suggest that the two populations in the green valley have followed distinct evolutionary pathways. Simulations indicate that while fast-rotating galaxies can continuously gain angular momentum through the accretion of infalling gas, slow-rotating galaxies experience more frequent mergers that efficiently redistribute and diminish their angular momentum. Our findings further suggest that such mergers are either associated with or directly induce strong outflows in slow-rotating galaxies across the entire stellar mass range. These outflows remove metals from the galaxy, contributing to the stellar metallicity deficiency observed in slow-rotating galaxies. However, the impact of the proposed merger events on star formation history appears to depend strongly on stellar mass.

    \item Summarizing evidence from our model-based analysis, together with results from semi-analytic models and cosmological simulations, we suggest the following picture. In high-mass galaxies, mergers experienced by the slower population may trigger strong AGN and star formation activity. The resulting feedback can rapidly deplete the galaxy’s gas reservoir and suppress further gas accretion. Consequently, massive slow-rotating galaxies quench their star formation much more quickly than their fast-rotating counterparts. In contrast, the evolution of low-mass galaxies is likely governed predominantly by supernova feedback. While capable of driving strong outflows, supernova feedback alone is generally insufficient to halt gas infall completely. As a result, low-mass fast- and slow-rotating galaxies exhibit similar gas infall timescales and star formation histories, despite their differences in chemical composition.

    \item Alternative mechanisms such as environmentally driven evolution or differences in assembly history may also contribute to the observational trends presented in this work. However, given the limitations of both our sample and the modelling framework, we do not find definitive evidence that clearly supports or rules out these possibilities.

\end{itemize}

The combined analysis of stellar and gas-phase properties is key to uncovering the evolutionary histories of galaxies. The framework of semi-analytic spectral fitting offers more than a convenient tool for reproducing observed spectra and gas-phase metallicities; when coupled with a physically motivated chemical-evolution prescription, it provides a means to connect spectral constraints to the physical processes that regulate star formation and chemical enrichment. It further provides a natural bridge between theoretical and simulation-based predictions and observational data. The ability of this approach to yield a coherent evolutionary scenario, together with the consistency between the model-based results and independent lines of evidence, lends support to the plausibility of the proposed interpretation. The consistency between our green valley results and trends seen in quenched early-type galaxies, despite slightly different selection criteria, supports a coherent kinematic evolution framework, while the precise evolutionary paths remain to be clarified by future observations.  This type of analysis will also help reveal how galaxy evolution depends on environment \citep{Zhou2022environment}, offering a powerful avenue to connect internal processes with large-scale structure. Forthcoming large spectroscopic surveys providing high-quality spectra of individual galaxies such as WEAVE-StePS \citep{Iovino2023} and, on a longer timescale, surveys utilizing IFU capabilities such as WST \citep{Bacon2024}, will target galaxies at higher redshifts, enabling further tests of the scenario proposed here and providing new insights into the interplay between star formation history, chemical evolution, and galaxy's dynamical status.

\label{sec:summary}

\begin{acknowledgements}
The authors would like to thank the anonymous referee for their constructive comments, which have significantly improved this work. S.Z., A.I., M.L. and F.L.B. acknowledge financial support from INAF Large Grant 2022, FFO 1.05.01.86.16. F.L.B. acknowledges support from INAF minigrant 1.05.23.04.01. 

The project that gave rise to these results received the support of a fellowship from the “la Caixa” Foundation (ID 100010434). The fellowship code is LCF/BQ/PR24/12050015. L.C. acknowledges support from grants PID2022-139567NB-I00 and PIB2021-127718NB-I00 funded by the Spanish Ministry of Science and Innovation/State Agency of Research  MCIN/AEI/10.13039/501100011033 and by “ERDF A way of making Europe”. 

Funding for the Sloan Digital Sky Survey IV has been provided by the Alfred P. 
Sloan Foundation, the U.S. Department of Energy Office of Science, and the Participating Institutions. 
SDSS-IV acknowledges support and resources from the Center for High-Performance Computing at 
the University of Utah. The SDSS web site is www.sdss.org.

SDSS-IV is managed by the Astrophysical Research Consortium for the Participating Institutions of the SDSS Collaboration including the Brazilian Participation Group, the Carnegie Institution for Science, Carnegie Mellon University, the Chilean Participation Group, the French Participation Group, Harvard-Smithsonian Center for Astrophysics, Instituto de Astrof\'isica de Canarias, The Johns Hopkins University, Kavli Institute for the Physics and Mathematics of the Universe (IPMU) / University of Tokyo, Lawrence Berkeley National Laboratory, Leibniz Institut f\"ur Astrophysik Potsdam (AIP), Max-Planck-Institut f\"ur Astronomie (MPIA Heidelberg), Max-Planck-Institut f\"ur Astrophysik (MPA Garching), Max-Planck-Institut f\"ur Extraterrestrische Physik (MPE), National Astronomical Observatories of China, New Mexico State University, New York University, University of Notre Dame, Observat\'ario Nacional / MCTI, The Ohio State University, Pennsylvania State University, Shanghai Astronomical Observatory, United Kingdom Participation Group, Universidad Nacional Aut\'onoma de M\'exico, University of Arizona, University of Colorado Boulder, University of Oxford, University of Portsmouth, University of Utah, University of Virginia, University of Washington, University of Wisconsin, Vanderbilt University, and Yale University.

\end{acknowledgements}

\bibliographystyle{aa}
\bibliography{szhou} 

\end{document}